\documentclass[aps,prl,reprint,superscriptaddress,amsmath,amssymb,preprintnumbers,showpacs]{revtex4-1}
\usepackage[normalem]{ulem}

\usepackage{enumerate}
\usepackage{graphicx}
\usepackage{color}
\usepackage{float} 
\usepackage{xspace}
\usepackage{bm} 
\graphicspath{{./}}

\newcommand{\ket}[1]{\bigl| #1 \bigr>} 
\newcommand{\bra}[1]{\bigl< #1 \bigr|} 
\newcommand{\abs}[1]{\left| #1 \right|} 

\newcommand{\floqel}[1]{\xi_{E_\text{e}}}
\newcommand{\floqnuc}[1]{\xi_{E_\text{N}}}

\renewcommand{\vec}[1]{{\mathbf{\bm{#1}}}}

\newcommand{\hBN}{h-BN}
\newcommand{\ehPE}{eh-PER}
\newcommand{\berryc}{\vec{\mathcal{A}}}

\newcommand{\com}{\vec{\mathcal{D}}}
\newcommand{\width}{\sigma}
\newcommand{\pol}{\mu}

\begin{document}
\title{Imperfect Recollisions in High-Harmonic Generation in Solids}
\author{Lun \surname{Yue}}
\email{lun\_yue@msn.com}
\author{Mette B. \surname{Gaarde}}
\email{gaarde@phys.lsu.edu}
\affiliation{Department of Physics and Astronomy, Louisiana State University, Baton Rouge, Louisiana 70803-4001, USA}
\date{\today} 

\begin{abstract}
  We theoretically investigate high-harmonic generation in hexagonal boron nitride with linearly polarized laser pulses. We show that imperfect recollisions between electron-hole pairs in the crystal give rise to an electron-hole-pair polarization energy that leads to a double-peak structure in the subcycle emission profiles. An extended recollision model (ERM) is developed that allows for such imperfect recollisions, as well as effects related to Berry connections, Berry curvatures, and transition-dipole phases. The ERM illuminates the distinct spectrotemporal characteristics of harmonics emitted parallel and perpendicularly to the laser polarization direction. 
  Imperfect recollisions are a general phenomenon and a manifestation of the spatially delocalized nature of the real-space wave packet, they arise naturally in systems with large Berry curvatures, or in any system driven by elliptically polarized light. 
\end{abstract}

\maketitle

The last decade has seen the emergence of high-harmonic generation (HHG) in solids \cite{Ghimire2011, Vampa2015Nat, You2016, Ndabashimiye2016, Garg2016, Wang2017} as a promising and compact ultrafast light source, as well as a potential tool to reconstruct crystal band structures \cite{Vampa2015b}, measure Berry curvatures \cite{Liu2017, Luu2018}, and probe topological phase transitions \cite{Bauer2018, Silva2019, Chacon2019}.
Complementing experimental progress, a number of theoretical studies have explored HHG in solids either in terms of reciprocal-space dynamics within the band structure \cite{Ghimire2011, Kemper2013, Vampa2014, Schubert2014, Higuchi2014, Luu2015, Wu2015, Hohenleutner2015, Garg2016, Tamaya2016, Wang2017, Garg2018, Li2019, Navarette2019}, which contains both intra- and interband contributions, or in terms of real-space particle-like dynamics in the crystal \cite{Vampa2014, Vampa2015, Yu2019, Vampa2015Nat, You2016, Ikemachi2017, Tancogne-Dejean2017, Tancogne-Dejean2018, Yoshikawa2019}.

\begin{figure}
  \centering
  \includegraphics[width=0.48\textwidth, clip, trim=0 0cm 0 0cm]{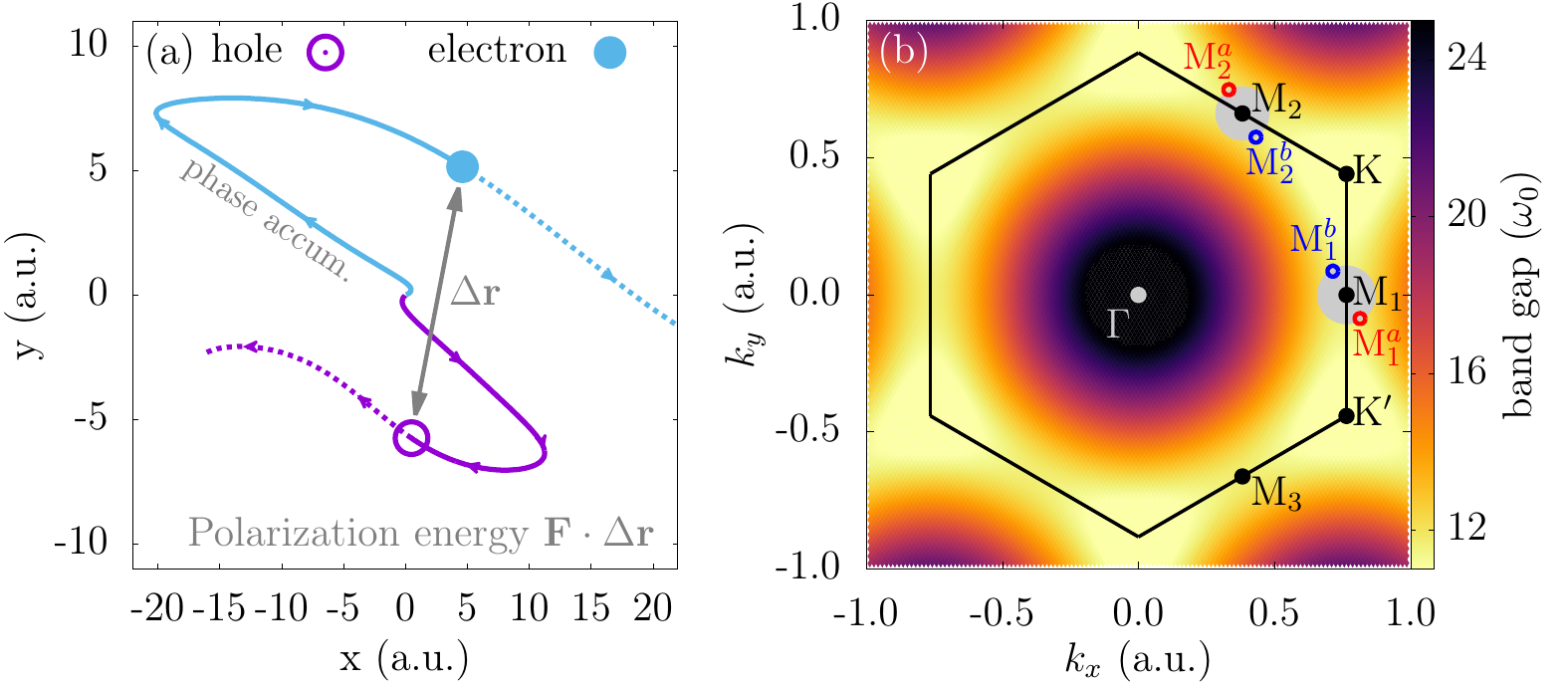}
  \caption{(a) An imperfect recollision in which the electron and hole centers do not exactly overlap, leading to an electron-hole-pair polarization energy upon recollision. In the ERM, we also keep track of the phase accumulated along the trajectories. (b) Band gap energy (in units of $\omega_0 = 0.0285$) of monolayer \hBN{} with annotated points of interest. The gray discs around $\text{M}_1$ and $\text{M}_2$ each has radius 0.1.}
  \label{fig:sketch}
\end{figure}

The semi-classical recollision model, extended to solids \cite{Vampa2014, Vampa2015} from the gas-phase three-step model of strong-field interactions \cite{Corkum1993, Lewenstein1994}, has in particular provided an intuitive understanding of the 
interband contribution to HHG in solids: 
in each laser half-cycle, an electron tunnels from the valence band to the conduction band and leaves behind a hole;
the laser field spatially drives the electron and hole according to their respective band-structure dispersions; when the electron and hole recollide, a high-energy photon is emitted with energy corresponding to the instantaneous band gap.
At the recollision step, the assumption has been that the electron and hole reencounter each other exactly at the same spatial position. However, since Bloch waves are spatially delocalized, it is possible for the electron and hole wavepackets to spatially overlap even when their centers do not [Fig.~\ref{fig:sketch}(a)], with the implication that an electron-hole-pair polarization energy at recollision (\ehPE) will contribute to the emitted photon energy. Also, the original recollision model does not discern between the parallel and perpendicularly polarized currents, and the omission of ubiquitous solid-state properties such as the Berry connection, Berry curvatures, and transition dipole phases (TDPs) are generally not well-justified. For instance, a recent numerical work \cite{Jiang2018} highlighted the importance of the TDPs on the generation of even-order harmonics in ZnO. It is thus desirable to construct a framework that is able to include all these afore-mentioned concepts.

In this Letter, we consider how imperfect recollisions manifest themselves in the HHG process in solids, and provide an extended recollision model (ERM) that naturally includes this effect, as well as other properties such as the Berry connections and the TDPs. As a concrete example, we study HHG in monolayer hexagonal boron nitride (\hBN{}) driven by infrared pulses linearly polarized in the crystal plane, and we compare the ERM results to solutions of the semiconductor Bloch equations (SBEs) \cite{Golde2008, Kira2012}. In addition to the recent intense interest in HHG from two-dimensional materials \cite{Yoshikawa2017, Taucer2017, Liu2017, Hafez2018, Tancogne-Dejean2018, Silva2019b, Mrudul2019}, \hBN{} is interesting due to its lack of an inversion center which leads to non-zero Berry connections and TDPs. Also, pulse propagation effects \cite{Floss2018, Lu2019} can be neglected in monolayer materials. We find that when the driving laser is polarized along the $\Gamma-\text{K}$ direction, the \ehPE{} manifests itself in the time-frequency spectrograms of the parallel-emitted odd-order harmonics as a double-peak structure. Within the ERM this is explained as two sets of trajectories launched each half-cycle, from different $\vec{k}$-points in the Brillouin zone (BZ). We show that the quantum interference between multiple $\vec{k}$-point contributions gives rise to different spectrotemporal characteristics of harmonic emission parallel or perpendicular to the laser polarization direction (LPD), when the Berry connections and the TDPs are included.
We find that the imperfect recollisions involve tens of a.u. separations at the time of recollision, and that this is consistent with the size of the delocalized quantum wave packet. 
The formulation of the ERM and in particular the inclusion of the \ehPE{} provides new insights into the HHG process in solids and could potentially stimulate new experiments.

\begin{figure}
  \centering
  \includegraphics[width=0.48\textwidth, clip, trim=0 0cm 0 0cm]{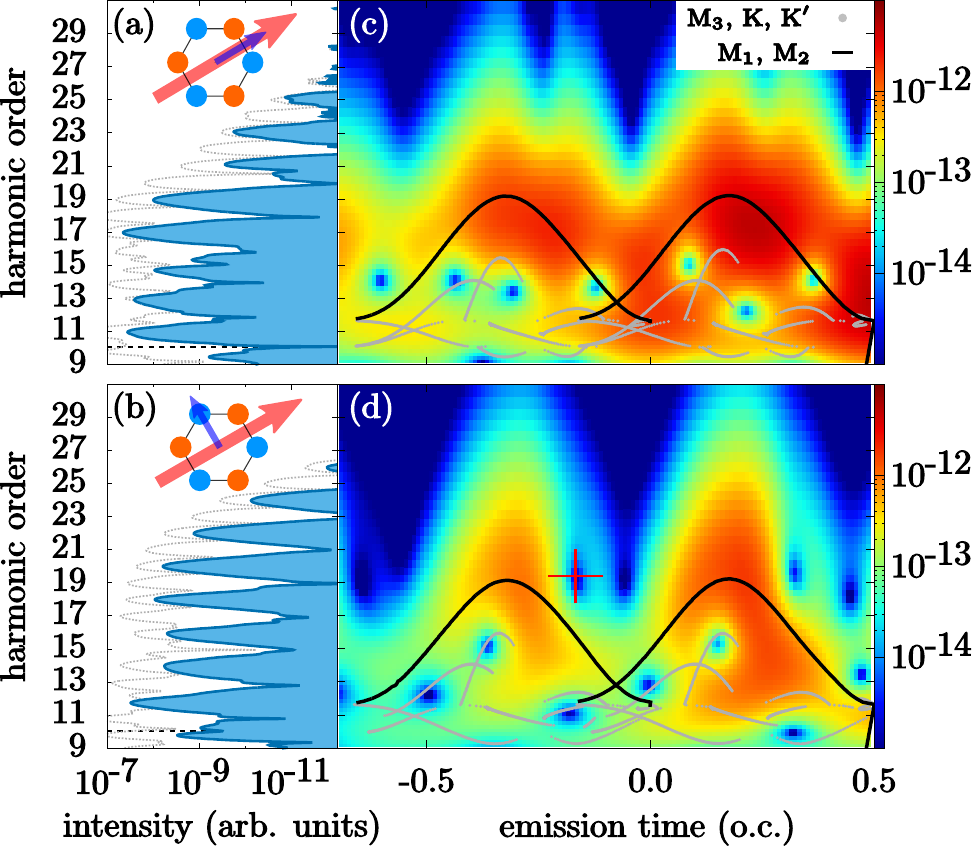}
  \caption{HHG spectra (a,b) and time-frequency profiles (c,d) for LPD along the $\Gamma-\text{K}$ direction, and high harmonics polarized parallel (a,c) and perpendicular (b,d) to the LPD. In (a) and (b), the filled curve results from the full SBE calculation, the dotted line neglects the Berry connections, and the dashed horizontal line outlines the minimal band gap. In (c) and (d), the color map is the quantum result; the lines and dots are semiclassical results, with tunneling from different symmetry points in the BZ [see Fig.~\ref{fig:sketch}(b)]. The red cross in (d) marks a minimum in the color map.}
  \label{fig:hhgcwt}
\end{figure}

We start by solving the SBEs for monolayer \hBN{} with inclusion of one valence and one conduction band, using a dephasing time $T_2=5\;\text{fs}$. The band structure of monolayer \hBN{} [Fig.~\ref{fig:sketch}(b)] is obtained by the pseudopotential method of Ref.~\cite{Taghizadeh2017} and is given in the Supplemental Material (SM) \cite{suppmat2020_imperf_recol}. The Berry connections and TDPs are calculated numerically in the twisted parallel transport gauge \cite{Vanderbilt2018}, ensuring continuity and BZ-periodicity. For the laser parameters in this paper, the HHG spectrum is converged with respect to increasing the number of bands. 
The filled curves in Figs.~\ref{fig:hhgcwt}(a) and \ref{fig:hhgcwt}(b) show the high-harmonic spectrum of monolayer \hBN{} driven by a 1600 nm, 35 fs laser pulse with a peak intensity of 3.5 $\text{TW/cm}^2$. Harmonics up to 25th order are generated, with purely odd (even) orders generated parallel (perpendicular) to the LPD.
The dotted lines in Figs.~\ref{fig:hhgcwt}(a) and \ref{fig:hhgcwt}(b) show that inclusion of only the TDPs without the Berry connections, as e.g. was done in \cite{Jiang2018} to show the emergence of even harmonics in ZnO, would produce erroneous spectra.

Figures~\ref{fig:hhgcwt}(c) and \ref{fig:hhgcwt}(d) show the sub-cycle time-frequency profiles for the emission with parallel and perpendicular polarization, respectively, and illustrate a key result of this paper: The parallel emission profile clearly exhibits a broad, double-peaked structure, whereas the perpendicular profile is single-peaked and much narrower, and in fact exhibits a pronounced minimum at the position corresponding to the second peak in the parallel emission.

To get a clear physical understanding of the emission dynamics observed in Fig.~\ref{fig:hhgcwt}, we develop an extended version of the recollision model for HHG in solids \cite{Vampa2014, Vampa2015}.
We focus on the interband dynamics which strongly dominates the emission above the band gap. 
Atomic units are used throughout this work unless indicated otherwise.
In the two-band approximation and assuming a small population in the conduction band, the interband spectrum is $\vec{j}(\omega) = \int_{-\infty}^{\infty} dt e^{i\omega t} \vec{j}(t)$, where the current components parallel and perpendicular ($\pol=\|, \perp$) to the LPD are 
\begin{equation}
  \label{eq:theory_sfa_1}
  \begin{aligned}
    j_\pol(t)
    & = \sum_{\vec{k}} R^{\vec{k}}_\pol  \int^t
     T^{\vec{\kappa}(s)}
    e^{-iS_\pol(\vec{k}, t, s)} ds
    + \text{c.c.}
  \end{aligned}
\end{equation}
with $T^{\vec{\kappa}(s)}=F(s) |{d}_{ \|}^{\vec{\kappa}(s)}|$ the transition matrix element, $R^{\vec{k}}_\pol =  \omega_{g}^{\vec{k}}| {d}_{\pol}^{\vec{k}} |$ the recombination dipole, $\omega_g^{\vec{k}} = E_c^{\vec{k}} - E_v^{\vec{k}}$ the band gap, $\vec{F}(t) = F(t) \vec{\hat{e}}_{\|}$ the electric field with $\vec{\hat{e}}_{\|}$ the LPD, $\vec{d}^{\vec{k}} = i\bra{u_{c}^{\vec{k}}} \nabla_{\vec{k}}\ket{u_{v}^{\vec{k}}}$ the coupling matrix elements with $u_m^{\vec{k}}$ the periodic part of the Bloch function, and $\vec{\kappa}(\tau) = \vec{k} - \vec{A}(t) + \vec{A}(\tau)$ with $\vec{A}$ the vector potential satisfying $- d\vec{A}/dt = \vec{F}$.
The accumulated phase in Eq.~\eqref{eq:theory_sfa_1} is (dephasing ignored)
\begin{equation}
  \label{eq:theory_sfa_2}
  \begin{aligned}
    S_\pol(\vec{k}, t, s)
    =& \int_s^t \left[ \omega_g^{\vec{\kappa}(\tau)} + \vec{F}(\tau)\cdot\Delta\berryc^{\vec{\kappa}(\tau)} \right] d\tau \\
    & + \alpha^{\vec{k}}_\pol - \alpha^{\vec{\kappa}(s)}_{\|} 
  \end{aligned}
\end{equation}
with $\Delta \berryc^{\vec{k}} = \berryc_c^{\vec{k}} - \berryc_v^{\vec{k}}$ where $\berryc_n^{\vec{k}} = i\bra{u_n^{\vec{k}}} \nabla_{\vec{k}}\ket{u_n^{\vec{k}}}$ are the Berry connections, and $\alpha^{\vec{k}}_\pol=-\arg({d}_\pol^{\vec{k}})$ the transition dipole phases (TDPs).
The saddle point conditions for $S_\pol(\vec{k},t,s)-\omega t$ are
\begin{subequations}
  \label{eq:theory_sfa_3}
  \begin{align}
    \omega_g^{\vec{\kappa}(s)} + \vec{F}(s) \cdot \com^{\vec{\kappa}(s)}_\parallel
    & = 0, \label{eq:theory_sfa_3a}
    \\
    \Delta \vec{R}_\pol \equiv \Delta\vec{r} - \com^{\vec{k}}_\pol + \com^{\vec{\kappa}(s)}_\parallel
    & = 0, \label{eq:theory_sfa_3b}
    \\
    \omega_g^{\vec{k}}
      + \vec{F}(t) \cdot \left[ \com^{\vec{\kappa}(s)}_\parallel + \Delta\vec{r}  \right]
    & = \omega, \label{eq:theory_sfa_3c}
  \end{align}
\end{subequations}
with $\com^{\vec{k}}_\pol \equiv \Delta \berryc^{\vec{k}} - \nabla_{\vec{k}}\alpha_\pol^{\vec{k}}$ and $\Delta \vec{r} \equiv \int_s^t \left[ \vec{v}_c^{\vec{\kappa}(\tau)} - \vec{v}_v^{\vec{\kappa}(\tau)} \right] d\tau$ where $\vec{v}_n^{\vec{\kappa}(\tau)} \equiv \nabla_{\vec{k}}E_n^{\vec{\kappa}(\tau)} + \vec{F}(\tau) \times \vec{\Omega}_n^{\vec{\kappa}(\tau)}$ is the velocity including the Berry curvature $\vec{\Omega}_n^{\vec{k}} \equiv \nabla_{\vec{k}} \times \berryc_n^{\vec{k}}$. Semiclassically, Eqs.~\eqref{eq:theory_sfa_3a}-\eqref{eq:theory_sfa_3c} are interpreted in terms of the three steps in the recollision model: at time $s$, an electron tunnels from the valence to the conduction band with crystal momentum $\vec{\kappa}(s)$; the newly created electron-hole pair is accelerated by the laser and recollides at time $t$ with final crystal momentum $\vec{k}$ and relative distance $\Delta\vec{r}$; at recollision, a high-energy photon with energy $\omega$ is released. Our ERM
\footnote{Note that similar equations appear in \cite{Li2019phase}, but without inclusion of the electron-hole polarization energy. They were applied to a model one-dimensional system where the Berry connections can naturally chosen to be zero. We stress that imperfect recollisions can only occur in systems with dimensions higher than one.}
extends previous works by including (i) laser-dressing of the bands with $\vec{F}\cdot\com_\|^{\kappa(s)}$; (ii) the Berry curvature contribution to the velocity of the trajectories; (iii) the \ehPE{} given by $\vec{F}\cdot \Delta\vec{r}$, due to the nonzero recollision distance $\Delta \vec{r}\ne \vec{0}$. In addition, we keep track of the accumulated phase of each trajectory. For \hBN{} and the field parameters used here, points (i) and (ii) are of minor importance \footnote{This is consistent with the discussion of laser dressing in Ref.~\cite{Tamaya2016}.}. In all our ERM calculations, a continuous-wave laser is employed, with tunneling times $s\in[-T, 0]$ where $T$ is the period. Returning trajectories with $\abs{\Delta\vec{R}_\pol}<R_{\text{0}}$ are assumed to have recollided, with $R_0\equiv 30$ unless indicated otherwise.


\begin{figure}
  \centering
  \includegraphics[width=0.48\textwidth, clip, trim=0 0cm 0 0cm]{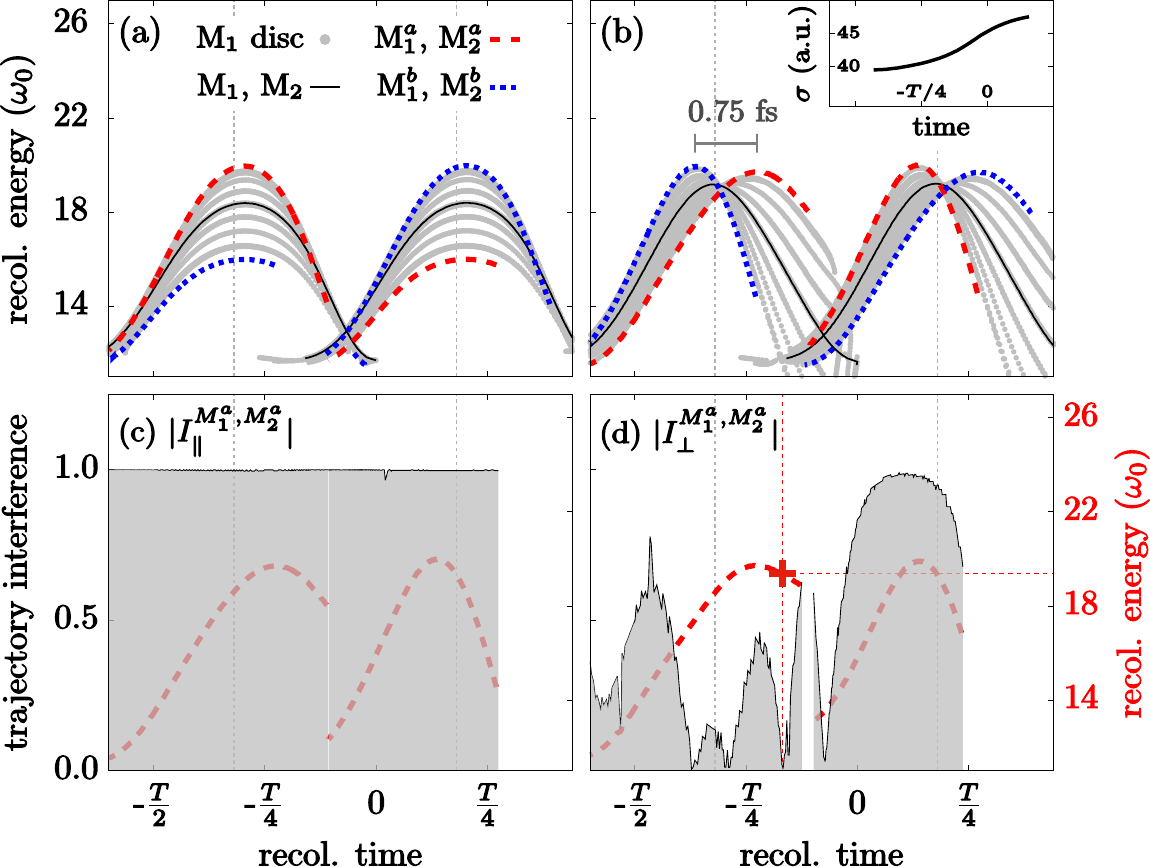}
  \caption{Semiclassical recollision energies versus recollision times (a) without and (b) with inclusion of the \ehPE{} $\vec{F}\cdot \Delta \vec{r}$ in Eq.~\eqref{eq:theory_sfa_3c}, for LPD along $\Gamma-\text{K}$. The different curves correspond to different tunneling sites in the BZ [Fig.~\ref{fig:sketch}(b)]. The vertical dotted lines mark the peaks of $\text{M}_1$ and $\text{M}_2$ recollision energies.
    The inset in (b) shows the spatial spread corresponding to the trajectory that tunnels at $\text{M}_1$ with the highest recollision energy.
    The shaded curves in (c,d) show the quantum interferences between the $\text{M}_1^a$ and $\text{M}_2^a$ emissions, for the (c) parallel and (d) perpendicular directions (see text). The recollision energies (right y-axis) for $\text{M}_1^a$ and $\text{M}_2^a$ trajectories are redrawn in red dashed lines. In (d), the red cross indicates the same interference minimum as in Fig.~\ref{fig:hhgcwt}(d).} 
  \label{fig:stark}
\end{figure}

In Figs.~\ref{fig:hhgcwt}(c) and \ref{fig:hhgcwt}(d), the semiclassical recollision energies are plotted for different initial crystal momenta $\vec{\kappa}(s)$ corresponding to different symmetry points in the BZ [Fig.~\ref{fig:sketch}(b)]. The $\text{M}_1$ and $\text{M}_2$ symmetry points are seen here to be responsible for the emitted high-order harmonics with orders $\gtrsim 16$. This is consistent with the density of states diverging at the $\text{M}$ points in \hBN{} \cite{Hipolito2016} (van Hove singularities \cite{vanHove1953}), and strong HHG emissions are expected at such points \cite{Uzan2018, Yoshikawa2019}.
The recollision energy only peaks once during each half-cycle (with each energy below the maximum being emitted twice, resulting from a "short" and "long" trajectory, respectively, similar to gas-phase HHG), which differs from the double peak structure in the quantum result of Fig.~\ref{fig:hhgcwt}(c).
 In addition, the recollision energies for the perpendicular harmonics are almost identical to those of the parallel case and do not reproduce the narrow SBE time profile [Fig.~\ref{fig:hhgcwt}(d)].

 The imperfect recollisions, as we now demonstrate, are responsible for the double-peak structure in Fig.~\ref{fig:hhgcwt}(c). Indeed, when we take into account \ehPE{} and all crystal momenta in a disc of radius 0.1 around the $\text{M}_1$ symmetry point in the BZ [henceforth referred to as the $\text{M}_1$ disc, see Fig.~\ref{fig:sketch}(b)], we recover the double peak structure in the semiclassical recollision model, shown in Fig.~\ref{fig:stark}(b) by the gray dots. The time-delay between the two emissions is  $\sim0.75$ fs, with the first emission slightly higher in energy, in agreement with the quantum result of Fig.~\ref{fig:hhgcwt}(c).
When the \ehPE s are neglected in Fig.~\ref{fig:stark}(a), the $\text{M}_1$ disc emits the highest-order harmonics at the same time, and no double-peak structure is observed. 
To better understand the role of the \ehPE s, we choose to consider two pairs of representative $\vec{k}$-points on the periphery of the $\text{M}_1$ and $\text{M}_2$ discs [red and blue circles in Fig.~\ref{fig:sketch}(b)]. As seen in Figs.~\ref{fig:stark}(a) and \ref{fig:stark}(b), both the recollision times and the recollision energies are modified when taking into account the \ehPE{}s: $\text{M}_1^a$ and $\text{M}_2^a$ ($\text{M}_1^b$ and $\text{M}_2^b$) recollide later (earlier with higher energy) during the first half-cycle, and earlier with higher energy (later) during the second half-cycle. Henceforth, we will refer to the two peaks during each half-cycle as early and late emissions.
In Figs.~\ref{fig:stark}(a) and \ref{fig:stark}(b), we have only considered the $\pol=\parallel$ case; for $\pol=\perp$, the result is similar, i.e. with a double-peak structure in the semiclassical recollision energies when including the \ehPE.

Note that our choice $R_0=30$ is several times greater than the \hBN{} lattice constant of 4.7, stressing the delocalized character of the spatial recollision process.
When decreasing $R_0$, the maximum recollision times of the $\text{M}_1^a$ and $\text{M}_1^b$ curves in Fig.~\ref{fig:stark}(b) become shorter, with the peak structures, and thereby all ``long'' trajectories, disappearing around $R_0\sim 20$ (see SM). For the electron and hole wave packets to spatially overlap at recollision time, the good agreement between our semiclassical and quantum results [Fig.~\ref{fig:stark}(b) and Fig.~\ref{fig:hhgcwt}(c)] thus suggests that the {\it quantum wave packet} has a minimum spread of $\sim 30$.
We have further estimated the quantum spatial spread by explicitly constructing a real-space wave packet during time propagation, after placing a $\vec{k}$-space wave packet on the conduction band (using the Houston-state basis \cite{Houston1940, Krieger1986}) with a $\vec{k}$-width estimated by tunneling (see SM for more details). 
For the trajectory that tunnels at the $\text{M}_1$ point and has the highest recollision energy, the spatial spread $\width$ of the corresponding wave packet is shown in the inset of Fig.~\ref{fig:stark}(b), where $\width$ is seen to increase from 39 to 48 between tunnel and recollision. These $\width$ values are fully consistent with our choice of $R_0$ and the previous discussion. The double-peak structure is also robust with respect to the choice of $T_2$, even up to 20 fs (see SM). This is consistent with our previous comments: from the time of tunneling, the electron and hole wave packets are driven apart spatially, allowing for recollision primarily during the first optical cycle of the pulse. 

\begin{figure}
  \centering
  \includegraphics[width=0.48\textwidth, clip, trim=0 0cm 0 0cm]{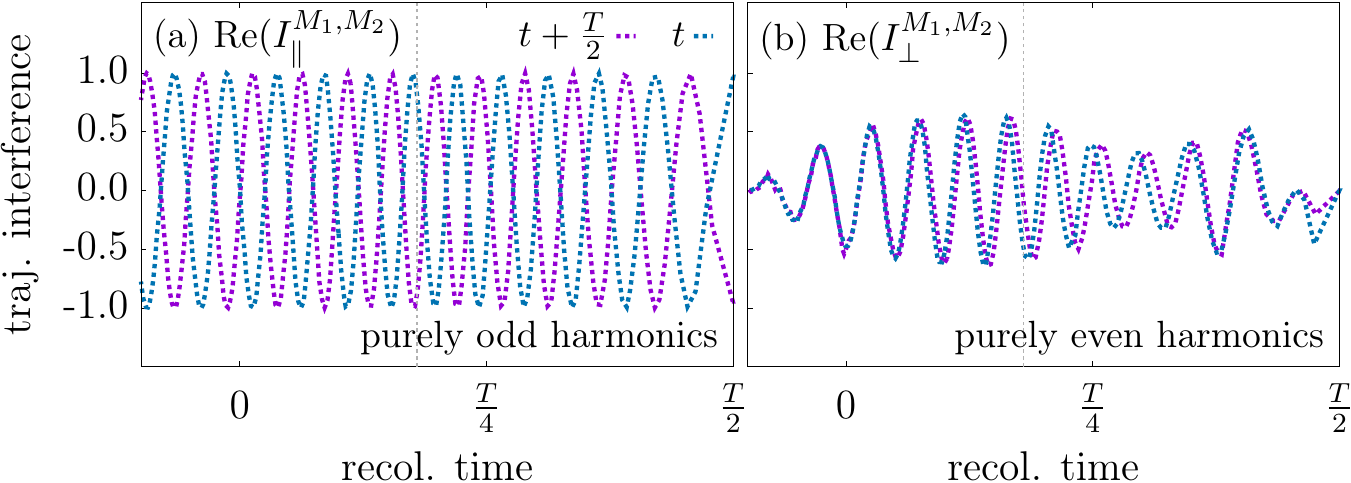} 
  \caption{Quantum interferences between the $\text{M}_1$ and $\text{M}_2$ emissions from consecutive half-cycles, for the (a) parallel and (b) perpendicular directions. The LPD is $\Gamma-\text{K}$.}
  \label{fig:phase}
\end{figure}

If the recollision model predicts a double-peak structure for both the parallel and perpendicular time-profiles, why do we not see this in the quantum result of Fig.~\ref{fig:hhgcwt}(d)? The answer lies in the quantum phase information of the trajectories. If two trajectories tunnel from two different $\vec{k}$-points in the BZ, $\text{P}_1$ and $\text{P}_2$, and have the same recollision energy at the same recollision time $t$, we can coherently add them as
\begin{equation}
  \label{eq:theory_sfa_4}
  I_\pol^{\text{P}_1\text{P}_2}(t) = \frac{1}{2} \left[e^{S^{\text{P}_1}_\pol(t)}+e^{S^{\text{P}_2}_\pol(t)}\right],
\end{equation}
with $S_\pol^{\text{P}_1}(t)$ and $S_\pol^{\text{P}_2}(t)$ the accumulated phases [Eq.~\eqref{eq:theory_sfa_2}]. 
For the parallel harmonics in Fig.~\ref{fig:stark}(c), $|I_\parallel^{\text{M}^a_1\text{M}^a_2}(t)|=1$ for all $t$, indicating that the emission from $\text{M}_1^a$ and $\text{M}_2^a$ is completely in-phase. In contrast, for the perpendicular case in Fig.~\ref{fig:stark}(d), the late (early) emissions exhibit clear destructive (constructive) interferences. In addition, the pronounced minimum in the perpendicular time-frequency profile in Fig.~\ref{fig:hhgcwt}(d) is exactly reproduced by the interference minimum in Fig.~\ref{fig:stark}(d) shown by the red cross. Thus with our ERM, we can even explain quantitative details in the emission profiles. Note that the difference in the accumulated phases between the parallel and perpendicular harmonics can be traced back to the TDPs $\alpha_{\pol}^{\vec{k}}$ in Eq.~\eqref{eq:theory_sfa_2}, stressing their importance for HHG in solids. Similarly,
within our framework, the generation of purely odd (even) harmonics in the parallel (perpendicular) directions shown in Fig.~\ref{fig:hhgcwt}(a) [Fig.~\ref{fig:hhgcwt}(b)] is easily explained by considering interference between emissions originating in tunneling from the $\text{M}_1$ and $\text{M}_2$ sites, as illustrated in Fig.~\ref{fig:phase}. For the parallel polarization, the condition $I_\|^{\text{M}_1\text{M}_2}(t) = -I_\|^{\text{M}_1\text{M}_2}(t+\frac{T}{2})$ leads to purely odd harmonics, whereas for the perpendicular polarization, only even harmonics are generated due to $I_\perp^{\text{M}_1\text{M}_2}(t) = I_\perp^{\text{M}_1\text{M}_2}(t+\frac{T}{2})$.

\begin{figure}
  \centering
  \includegraphics[width=0.48\textwidth, clip, trim=0 0cm 0 0cm]{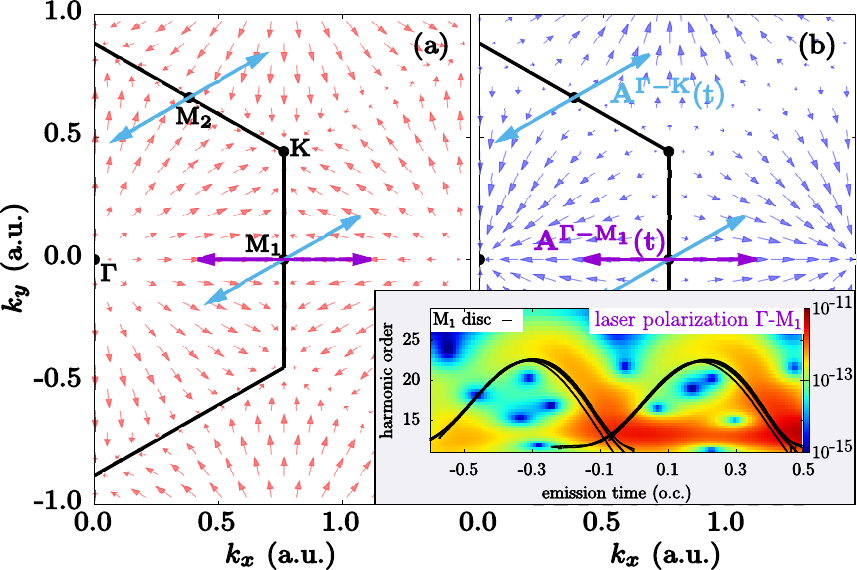}
  \caption{Relevance of the \ehPE{} for linearly polarized drivers. The vector fields show the group velocities (without the anomalous term involving the Berry curvature) $\nabla_{\vec{k}}E_n^{\vec{k}}$ for (a) the hole and (b) the electron. For clarity, the vector-field size in (a) is scaled by 2 compared to (b).  The teal (purple) arrows sketch the excursion of the relevant electron-hole pairs in $\vec{k}$-space for a vector potential $\vec{A}(t)$ polarized along $\Gamma-\text{K}$ ($\Gamma-\text{M}_1$). Inset: the time-frequency profile and the semiclassical predictions, for a driver with $\Gamma-\text{M}_1$ polarization.}
  \label{fig:GM}
\end{figure}

Finally, we discuss the general situations in which imperfect recollisions should be important for HHG in solids. Clearly, the \ehPE{} contribution to the harmonic energy is ubiquitous in all solids where the electron and hole do not exactly spatially recollide. Obvious examples include HHG in solids with elliptical drivers, and systems with large Berry curvatures. For linear polarization, the general rule is the following: suppose an electron-hole-pair is created at a symmetry point $\text{S}_0$ and subsequently driven by a LPD along $\Gamma - \text{S}$, then the \ehPE{} will be important when $\text{S} \ne \text{S}_0$. This is illustrated in Fig.~\ref{fig:GM}, where for the $\Gamma-\text{K}$ polarization, the excursions of the electron-hole-pairs created at $\text{M}_1$ and $\text{M}_2$ in $\vec{k}$-space, $\vec{\kappa}(t)$ (teal arrows), are not along the group velocity vector fields $\nabla_{\vec{k}}E_n^{\vec{k}}$, yielding a nonzero \ehPE. In contrast, when the LPD is $\Gamma-\text{M}_1$, the $\vec{k}$-space excursion (purple arrows) are along the group velocities, such that the electron and hole travel along a straight line in real space, leading to perfect recollisions with $\Delta \vec{r}=\vec{0}$. Indeed, this is confirmed by both our quantum and semiclassical calculations in the inset of Fig.~\ref{fig:GM}, where no double peak is observed and all trajectories starting from a $\text{M}_1$ disc in the BZ have approximately the same emission times. We emphasize that the double-peak time profiles originate in emission from different $\vec{k}$-points in the BZ with similar emission times and we therefore do not expect them to be significantly altered by macroscopic effects such as phase matching or intensity averaging \cite{Floss2018}.
Indeed, we checked that the emission structures remain even when using $T_2=2\;\text{fs}$ in the SBEs, and are robust with respect to focal-volume averaging and spatial filtering in the far field \cite{Abadie2018}.

In conclusion, we have uncovered and characterized the effects of the imperfect recollisions for HHG in solids. In the example system of \hBN, they manifest in the time-frequency profiles as a
double peak structure in the sub-cycle emission profile of the parallel-polarized harmonics, which is absent in the perpendicular-polarized emission. In the process, we formulated an ERM for HHG in solids that captures the effects of the \ehPE, as well the implications of the Berry connections and TDPs.
We found that the spatial width of the electron and hole wave packets can be almost one order of magnitude larger than the lattice constant, allowing for the imperfect recollisions.
This suggests that the harmonic emission can probe the degree of spatial homogeneity of the periodic structure \cite{Huang2017, Almalki2018, Yu2019, Mrudul2019, Chinzei2020} as well as the temporal dephasing introduced by e.g. electron correlation.
We predict that \ehPE{} should be ubiquitous in a large range of extreme nonlinear phenomena of current interest, such as HHG with elliptical drivers \cite{Ghimire2011, You2016, Ndabashimiye2016, Tancogne-Dejean2017, Yoshikawa2017, Liu2017, Zurron-Cifuentes2019, Zhang2019} and systems with large Berry curvatures \cite{Xiao2010}, as well as high-order sideband generation \cite{Zaks2012, Langer2016, Banks2017, Langer2018}.
The identification of \ehPE{} and its effect on the harmonic emission profiles, as well as the formulation of the ERM, provides new insights into the HHG process in solids and could potentially stimulate new experiments. The potential experimental measurement of such subcycle emission dynamics could also give us information on where in the BZ the trajectories emanate, thus probing the dynamical band structures.
More generally, the characterization and understanding of harmonic emission times and spectral phases are very important for attosecond metrology in solids \cite{Vampa2015Nat, Hohenleutner2015, Garg2016, Garg2018, Lu2019}.


\acknowledgements
The authors acknowledge support from the National Science Foundation, under Grant No. PHY1713671. L. Y. greatfully acknowledges Francois Mauger and Kenneth J. Schafer for useful discussions.


\begin{thebibliography}{63}%
\makeatletter
\providecommand \@ifxundefined [1]{%
 \@ifx{#1\undefined}
}%
\providecommand \@ifnum [1]{%
 \ifnum #1\expandafter \@firstoftwo
 \else \expandafter \@secondoftwo
 \fi
}%
\providecommand \@ifx [1]{%
 \ifx #1\expandafter \@firstoftwo
 \else \expandafter \@secondoftwo
 \fi
}%
\providecommand \natexlab [1]{#1}%
\providecommand \enquote  [1]{``#1''}%
\providecommand \bibnamefont  [1]{#1}%
\providecommand \bibfnamefont [1]{#1}%
\providecommand \citenamefont [1]{#1}%
\providecommand \href@noop [0]{\@secondoftwo}%
\providecommand \href [0]{\begingroup \@sanitize@url \@href}%
\providecommand \@href[1]{\@@startlink{#1}\@@href}%
\providecommand \@@href[1]{\endgroup#1\@@endlink}%
\providecommand \@sanitize@url [0]{\catcode `\\12\catcode `\$12\catcode
  `\&12\catcode `\#12\catcode `\^12\catcode `\_12\catcode `\%12\relax}%
\providecommand \@@startlink[1]{}%
\providecommand \@@endlink[0]{}%
\providecommand \url  [0]{\begingroup\@sanitize@url \@url }%
\providecommand \@url [1]{\endgroup\@href {#1}{\urlprefix }}%
\providecommand \urlprefix  [0]{URL }%
\providecommand \Eprint [0]{\href }%
\providecommand \doibase [0]{http://dx.doi.org/}%
\providecommand \selectlanguage [0]{\@gobble}%
\providecommand \bibinfo  [0]{\@secondoftwo}%
\providecommand \bibfield  [0]{\@secondoftwo}%
\providecommand \translation [1]{[#1]}%
\providecommand \BibitemOpen [0]{}%
\providecommand \bibitemStop [0]{}%
\providecommand \bibitemNoStop [0]{.\EOS\space}%
\providecommand \EOS [0]{\spacefactor3000\relax}%
\providecommand \BibitemShut  [1]{\csname bibitem#1\endcsname}%
\let\auto@bib@innerbib\@empty
\bibitem [{\citenamefont {Ghimire}\ \emph {et~al.}(2011)\citenamefont
  {Ghimire}, \citenamefont {DiChiara}, \citenamefont {Sistrunk}, \citenamefont
  {Agostini}, \citenamefont {DiMauro},\ and\ \citenamefont
  {Reis}}]{Ghimire2011}%
  \BibitemOpen
  \bibfield  {author} {\bibinfo {author} {\bibfnamefont {S.}~\bibnamefont
  {Ghimire}}, \bibinfo {author} {\bibfnamefont {A.~D.}\ \bibnamefont
  {DiChiara}}, \bibinfo {author} {\bibfnamefont {E.}~\bibnamefont {Sistrunk}},
  \bibinfo {author} {\bibfnamefont {P.}~\bibnamefont {Agostini}}, \bibinfo
  {author} {\bibfnamefont {L.~F.}\ \bibnamefont {DiMauro}}, \ and\ \bibinfo
  {author} {\bibfnamefont {D.~A.}\ \bibnamefont {Reis}},\ }\bibfield  {title}
  {\enquote {\bibinfo {title} {Observation of high-order harmonic generation in
  a bulk crystal},}\ }\href {\doibase 10.1038/nphys1847} {\bibfield  {journal}
  {\bibinfo  {journal} {Nat. Phys.}\ }\textbf {\bibinfo {volume} {7}},\
  \bibinfo {pages} {138--141} (\bibinfo {year} {2011})}\BibitemShut {NoStop}%
\bibitem [{\citenamefont {Vampa}\ \emph
  {et~al.}(2015{\natexlab{a}})\citenamefont {Vampa}, \citenamefont {Hammond},
  \citenamefont {Thir{\'e}}, \citenamefont {Schmidt}, \citenamefont
  {L{\'e}gar{\'e}}, \citenamefont {McDonald}, \citenamefont {Brabec},\ and\
  \citenamefont {Corkum}}]{Vampa2015Nat}%
  \BibitemOpen
  \bibfield  {author} {\bibinfo {author} {\bibfnamefont {G.}~\bibnamefont
  {Vampa}}, \bibinfo {author} {\bibfnamefont {T.~J.}\ \bibnamefont {Hammond}},
  \bibinfo {author} {\bibfnamefont {N.}~\bibnamefont {Thir{\'e}}}, \bibinfo
  {author} {\bibfnamefont {B.~E.}\ \bibnamefont {Schmidt}}, \bibinfo {author}
  {\bibfnamefont {F.}~\bibnamefont {L{\'e}gar{\'e}}}, \bibinfo {author}
  {\bibfnamefont {C.~R.}\ \bibnamefont {McDonald}}, \bibinfo {author}
  {\bibfnamefont {T.}~\bibnamefont {Brabec}}, \ and\ \bibinfo {author}
  {\bibfnamefont {P.~B.}\ \bibnamefont {Corkum}},\ }\bibfield  {title}
  {\enquote {\bibinfo {title} {Linking high harmonics from gases and solids},}\
  }\href {https://doi.org/10.1038/nature14517} {\bibfield  {journal} {\bibinfo
  {journal} {Nature}\ }\textbf {\bibinfo {volume} {522}},\ \bibinfo {pages}
  {462 EP --} (\bibinfo {year} {2015}{\natexlab{a}})}\BibitemShut {NoStop}%
\bibitem [{\citenamefont {You}\ \emph {et~al.}(2016)\citenamefont {You},
  \citenamefont {Reis},\ and\ \citenamefont {Ghimire}}]{You2016}%
  \BibitemOpen
  \bibfield  {author} {\bibinfo {author} {\bibfnamefont {Y.~S.}\ \bibnamefont
  {You}}, \bibinfo {author} {\bibfnamefont {D.~A.}\ \bibnamefont {Reis}}, \
  and\ \bibinfo {author} {\bibfnamefont {S.}~\bibnamefont {Ghimire}},\
  }\bibfield  {title} {\enquote {\bibinfo {title} {Anisotropic high-harmonic
  generation in bulk crystals},}\ }\href {https://doi.org/10.1038/nphys3955}
  {\bibfield  {journal} {\bibinfo  {journal} {Nat. Phys.}\ }\textbf {\bibinfo
  {volume} {13}},\ \bibinfo {pages} {345 EP --} (\bibinfo {year}
  {2016})}\BibitemShut {NoStop}%
\bibitem [{\citenamefont {Ndabashimiye}\ \emph {et~al.}(2016)\citenamefont
  {Ndabashimiye}, \citenamefont {Ghimire}, \citenamefont {Wu}, \citenamefont
  {Browne}, \citenamefont {Schafer}, \citenamefont {Gaarde},\ and\
  \citenamefont {Reis}}]{Ndabashimiye2016}%
  \BibitemOpen
  \bibfield  {author} {\bibinfo {author} {\bibfnamefont {G.}~\bibnamefont
  {Ndabashimiye}}, \bibinfo {author} {\bibfnamefont {S.}~\bibnamefont
  {Ghimire}}, \bibinfo {author} {\bibfnamefont {M.}~\bibnamefont {Wu}},
  \bibinfo {author} {\bibfnamefont {D.~A.}\ \bibnamefont {Browne}}, \bibinfo
  {author} {\bibfnamefont {K.~J.}\ \bibnamefont {Schafer}}, \bibinfo {author}
  {\bibfnamefont {M.~B.}\ \bibnamefont {Gaarde}}, \ and\ \bibinfo {author}
  {\bibfnamefont {D.~A.}\ \bibnamefont {Reis}},\ }\bibfield  {title} {\enquote
  {\bibinfo {title} {Solid-state harmonics beyond the atomic limit},}\ }\href
  {https://doi.org/10.1038/nature17660} {\bibfield  {journal} {\bibinfo
  {journal} {Nature}\ }\textbf {\bibinfo {volume} {534}},\ \bibinfo {pages}
  {520 EP --} (\bibinfo {year} {2016})}\BibitemShut {NoStop}%
\bibitem [{\citenamefont {Garg}\ \emph {et~al.}(2016)\citenamefont {Garg},
  \citenamefont {Zhan}, \citenamefont {Luu}, \citenamefont {Lakhotia},
  \citenamefont {Klostermann}, \citenamefont {Guggenmos},\ and\ \citenamefont
  {Goulielmakis}}]{Garg2016}%
  \BibitemOpen
  \bibfield  {author} {\bibinfo {author} {\bibfnamefont {M.}~\bibnamefont
  {Garg}}, \bibinfo {author} {\bibfnamefont {M.}~\bibnamefont {Zhan}}, \bibinfo
  {author} {\bibfnamefont {T.~T.}\ \bibnamefont {Luu}}, \bibinfo {author}
  {\bibfnamefont {H.}~\bibnamefont {Lakhotia}}, \bibinfo {author}
  {\bibfnamefont {T.}~\bibnamefont {Klostermann}}, \bibinfo {author}
  {\bibfnamefont {A.}~\bibnamefont {Guggenmos}}, \ and\ \bibinfo {author}
  {\bibfnamefont {E.}~\bibnamefont {Goulielmakis}},\ }\bibfield  {title}
  {\enquote {\bibinfo {title} {Multi-petahertz electronic metrology},}\ }\href
  {https://doi.org/10.1038/nature19821} {\bibfield  {journal} {\bibinfo
  {journal} {Nature}\ }\textbf {\bibinfo {volume} {538}},\ \bibinfo {pages}
  {359 EP --} (\bibinfo {year} {2016})}\BibitemShut {NoStop}%
\bibitem [{\citenamefont {Wang}\ \emph {et~al.}(2017)\citenamefont {Wang},
  \citenamefont {Park}, \citenamefont {Lai}, \citenamefont {Xu}, \citenamefont
  {Blaga}, \citenamefont {Yang}, \citenamefont {Agostini},\ and\ \citenamefont
  {DiMauro}}]{Wang2017}%
  \BibitemOpen
  \bibfield  {author} {\bibinfo {author} {\bibfnamefont {Z.}~\bibnamefont
  {Wang}}, \bibinfo {author} {\bibfnamefont {H.}~\bibnamefont {Park}}, \bibinfo
  {author} {\bibfnamefont {Y.~H.}\ \bibnamefont {Lai}}, \bibinfo {author}
  {\bibfnamefont {J.}~\bibnamefont {Xu}}, \bibinfo {author} {\bibfnamefont
  {C.~I.}\ \bibnamefont {Blaga}}, \bibinfo {author} {\bibfnamefont
  {F.}~\bibnamefont {Yang}}, \bibinfo {author} {\bibfnamefont {P.}~\bibnamefont
  {Agostini}}, \ and\ \bibinfo {author} {\bibfnamefont {L.~F.}\ \bibnamefont
  {DiMauro}},\ }\bibfield  {title} {\enquote {\bibinfo {title} {The roles of
  photo-carrier doping and driving wavelength in high harmonic generation from
  a semiconductor},}\ }\href {\doibase 10.1038/s41467-017-01899-1} {\bibfield
  {journal} {\bibinfo  {journal} {Nat. Commun.}\ }\textbf {\bibinfo {volume}
  {8}},\ \bibinfo {pages} {1686} (\bibinfo {year} {2017})}\BibitemShut
  {NoStop}%
\bibitem [{\citenamefont {Vampa}\ \emph
  {et~al.}(2015{\natexlab{b}})\citenamefont {Vampa}, \citenamefont {Hammond},
  \citenamefont {Thir\'e}, \citenamefont {Schmidt}, \citenamefont {L\'egar\'e},
  \citenamefont {McDonald}, \citenamefont {Brabec}, \citenamefont {Klug},\ and\
  \citenamefont {Corkum}}]{Vampa2015b}%
  \BibitemOpen
  \bibfield  {author} {\bibinfo {author} {\bibfnamefont {G.}~\bibnamefont
  {Vampa}}, \bibinfo {author} {\bibfnamefont {T.~J.}\ \bibnamefont {Hammond}},
  \bibinfo {author} {\bibfnamefont {N.}~\bibnamefont {Thir\'e}}, \bibinfo
  {author} {\bibfnamefont {B.~E.}\ \bibnamefont {Schmidt}}, \bibinfo {author}
  {\bibfnamefont {F.}~\bibnamefont {L\'egar\'e}}, \bibinfo {author}
  {\bibfnamefont {C.~R.}\ \bibnamefont {McDonald}}, \bibinfo {author}
  {\bibfnamefont {T.}~\bibnamefont {Brabec}}, \bibinfo {author} {\bibfnamefont
  {D.~D.}\ \bibnamefont {Klug}}, \ and\ \bibinfo {author} {\bibfnamefont
  {P.~B.}\ \bibnamefont {Corkum}},\ }\bibfield  {title} {\enquote {\bibinfo
  {title} {All-optical reconstruction of crystal band structure},}\ }\href
  {\doibase 10.1103/PhysRevLett.115.193603} {\bibfield  {journal} {\bibinfo
  {journal} {Phys. Rev. Lett.}\ }\textbf {\bibinfo {volume} {115}},\ \bibinfo
  {pages} {193603} (\bibinfo {year} {2015}{\natexlab{b}})}\BibitemShut
  {NoStop}%
\bibitem [{\citenamefont {Liu}\ \emph {et~al.}(2017)\citenamefont {Liu},
  \citenamefont {Li}, \citenamefont {You}, \citenamefont {Ghimire},
  \citenamefont {Heinz},\ and\ \citenamefont {Reis}}]{Liu2017}%
  \BibitemOpen
  \bibfield  {author} {\bibinfo {author} {\bibfnamefont {H.}~\bibnamefont
  {Liu}}, \bibinfo {author} {\bibfnamefont {Y.}~\bibnamefont {Li}}, \bibinfo
  {author} {\bibfnamefont {Y.~S.}\ \bibnamefont {You}}, \bibinfo {author}
  {\bibfnamefont {S.}~\bibnamefont {Ghimire}}, \bibinfo {author} {\bibfnamefont
  {T.~F.}\ \bibnamefont {Heinz}}, \ and\ \bibinfo {author} {\bibfnamefont
  {D.~A.}\ \bibnamefont {Reis}},\ }\bibfield  {title} {\enquote {\bibinfo
  {title} {High-harmonic generation from an atomically thin semiconductor},}\
  }\href {https://doi.org/10.1038/nphys3946} {\bibfield  {journal} {\bibinfo
  {journal} {Nat. Phys.}\ }\textbf {\bibinfo {volume} {13}},\ \bibinfo {pages}
  {262 EP --} (\bibinfo {year} {2017})}\BibitemShut {NoStop}%
\bibitem [{\citenamefont {Luu}\ and\ \citenamefont
  {W{\"o}rner}(2018)}]{Luu2018}%
  \BibitemOpen
  \bibfield  {author} {\bibinfo {author} {\bibfnamefont {T.~T.}\ \bibnamefont
  {Luu}}\ and\ \bibinfo {author} {\bibfnamefont {H.~J.}\ \bibnamefont
  {W{\"o}rner}},\ }\bibfield  {title} {\enquote {\bibinfo {title} {Measurement
  of the berry curvature of solids using high-harmonic spectroscopy},}\ }\href
  {\doibase 10.1038/s41467-018-03397-4} {\bibfield  {journal} {\bibinfo
  {journal} {Nat. Commun.}\ }\textbf {\bibinfo {volume} {9}},\ \bibinfo {pages}
  {916} (\bibinfo {year} {2018})}\BibitemShut {NoStop}%
\bibitem [{\citenamefont {Bauer}\ and\ \citenamefont
  {Hansen}(2018)}]{Bauer2018}%
  \BibitemOpen
  \bibfield  {author} {\bibinfo {author} {\bibfnamefont {D.}~\bibnamefont
  {Bauer}}\ and\ \bibinfo {author} {\bibfnamefont {K.~K.}\ \bibnamefont
  {Hansen}},\ }\bibfield  {title} {\enquote {\bibinfo {title} {High-harmonic
  generation in solids with and without topological edge states},}\ }\href
  {\doibase 10.1103/PhysRevLett.120.177401} {\bibfield  {journal} {\bibinfo
  {journal} {Phys. Rev. Lett.}\ }\textbf {\bibinfo {volume} {120}},\ \bibinfo
  {pages} {177401} (\bibinfo {year} {2018})}\BibitemShut {NoStop}%
\bibitem [{\citenamefont {Silva}\ \emph
  {et~al.}(2019{\natexlab{a}})\citenamefont {Silva}, \citenamefont
  {Jim{\'e}nez-Gal{\'a}n}, \citenamefont {Amorim}, \citenamefont {Smirnova},\
  and\ \citenamefont {Ivanov}}]{Silva2019}%
  \BibitemOpen
  \bibfield  {author} {\bibinfo {author} {\bibfnamefont {R.~E.~F.}\
  \bibnamefont {Silva}}, \bibinfo {author} {\bibfnamefont {{\'A}.}~\bibnamefont
  {Jim{\'e}nez-Gal{\'a}n}}, \bibinfo {author} {\bibfnamefont {B.}~\bibnamefont
  {Amorim}}, \bibinfo {author} {\bibfnamefont {O.}~\bibnamefont {Smirnova}}, \
  and\ \bibinfo {author} {\bibfnamefont {M.}~\bibnamefont {Ivanov}},\
  }\bibfield  {title} {\enquote {\bibinfo {title} {Topological strong-field
  physics on sub-laser-cycle timescale},}\ }\href {\doibase
  10.1038/s41566-019-0516-1} {\bibfield  {journal} {\bibinfo  {journal} {Nat.
  Photonics}\ }\textbf {\bibinfo {volume} {13}},\ \bibinfo {pages} {849}
  (\bibinfo {year} {2019}{\natexlab{a}})}\BibitemShut {NoStop}%
\bibitem [{\citenamefont {Chac{\'o}n}\ \emph {et~al.}(2019)\citenamefont
  {Chac{\'o}n}, \citenamefont {Zhu}, \citenamefont {Kelly}, \citenamefont
  {Dauphin}, \citenamefont {Pisanty}, \citenamefont {Pic{\'o}n}, \citenamefont
  {Ticknor}, \citenamefont {Ciappina}, \citenamefont {Saxena},\ and\
  \citenamefont {Lewenstein}}]{Chacon2019}%
  \BibitemOpen
  \bibfield  {author} {\bibinfo {author} {\bibfnamefont {A.}~\bibnamefont
  {Chac{\'o}n}}, \bibinfo {author} {\bibfnamefont {W.}~\bibnamefont {Zhu}},
  \bibinfo {author} {\bibfnamefont {S.~P.}\ \bibnamefont {Kelly}}, \bibinfo
  {author} {\bibfnamefont {A.}~\bibnamefont {Dauphin}}, \bibinfo {author}
  {\bibfnamefont {E.}~\bibnamefont {Pisanty}}, \bibinfo {author} {\bibfnamefont
  {A.}~\bibnamefont {Pic{\'o}n}}, \bibinfo {author} {\bibfnamefont
  {C.}~\bibnamefont {Ticknor}}, \bibinfo {author} {\bibfnamefont {M.~F.}\
  \bibnamefont {Ciappina}}, \bibinfo {author} {\bibfnamefont {A.}~\bibnamefont
  {Saxena}}, \ and\ \bibinfo {author} {\bibfnamefont {M.}~\bibnamefont
  {Lewenstein}},\ }\bibfield  {title} {\enquote {\bibinfo {title} {Observing
  topological phase transitions with high harmonic generation},}\ }\href@noop
  {} {\bibfield  {journal} {\bibinfo  {journal} {arXiv:1807.01616v2}\ }
  (\bibinfo {year} {2019})}\BibitemShut {NoStop}%
\bibitem [{\citenamefont {Kemper}\ \emph {et~al.}(2013)\citenamefont {Kemper},
  \citenamefont {Moritz}, \citenamefont {Freericks},\ and\ \citenamefont
  {Devereaux}}]{Kemper2013}%
  \BibitemOpen
  \bibfield  {author} {\bibinfo {author} {\bibfnamefont {A.~F.}\ \bibnamefont
  {Kemper}}, \bibinfo {author} {\bibfnamefont {B.}~\bibnamefont {Moritz}},
  \bibinfo {author} {\bibfnamefont {J.~K.}\ \bibnamefont {Freericks}}, \ and\
  \bibinfo {author} {\bibfnamefont {T.~P.}\ \bibnamefont {Devereaux}},\
  }\bibfield  {title} {\enquote {\bibinfo {title} {Theoretical description of
  high-order harmonic generation in solids},}\ }\href {\doibase
  10.1088/1367-2630/15/2/023003} {\bibfield  {journal} {\bibinfo  {journal}
  {New Journal of Physics}\ }\textbf {\bibinfo {volume} {15}},\ \bibinfo
  {pages} {023003} (\bibinfo {year} {2013})}\BibitemShut {NoStop}%
\bibitem [{\citenamefont {Vampa}\ \emph {et~al.}(2014)\citenamefont {Vampa},
  \citenamefont {McDonald}, \citenamefont {Orlando}, \citenamefont {Klug},
  \citenamefont {Corkum},\ and\ \citenamefont {Brabec}}]{Vampa2014}%
  \BibitemOpen
  \bibfield  {author} {\bibinfo {author} {\bibfnamefont {G.}~\bibnamefont
  {Vampa}}, \bibinfo {author} {\bibfnamefont {C.~R.}\ \bibnamefont {McDonald}},
  \bibinfo {author} {\bibfnamefont {G.}~\bibnamefont {Orlando}}, \bibinfo
  {author} {\bibfnamefont {D.~D.}\ \bibnamefont {Klug}}, \bibinfo {author}
  {\bibfnamefont {P.~B.}\ \bibnamefont {Corkum}}, \ and\ \bibinfo {author}
  {\bibfnamefont {T.}~\bibnamefont {Brabec}},\ }\bibfield  {title} {\enquote
  {\bibinfo {title} {Theoretical analysis of high-harmonic generation in
  solids},}\ }\href {\doibase 10.1103/PhysRevLett.113.073901} {\bibfield
  {journal} {\bibinfo  {journal} {Phys. Rev. Lett.}\ }\textbf {\bibinfo
  {volume} {113}},\ \bibinfo {pages} {073901} (\bibinfo {year}
  {2014})}\BibitemShut {NoStop}%
\bibitem [{\citenamefont {Schubert}\ \emph {et~al.}(2014)\citenamefont
  {Schubert}, \citenamefont {Hohenleutner}, \citenamefont {Langer},
  \citenamefont {Urbanek}, \citenamefont {Lange}, \citenamefont {Huttner},
  \citenamefont {Golde}, \citenamefont {Meier}, \citenamefont {Kira},
  \citenamefont {Koch},\ and\ \citenamefont {Huber}}]{Schubert2014}%
  \BibitemOpen
  \bibfield  {author} {\bibinfo {author} {\bibfnamefont {O.}~\bibnamefont
  {Schubert}}, \bibinfo {author} {\bibfnamefont {M.}~\bibnamefont
  {Hohenleutner}}, \bibinfo {author} {\bibfnamefont {F.}~\bibnamefont
  {Langer}}, \bibinfo {author} {\bibfnamefont {B.}~\bibnamefont {Urbanek}},
  \bibinfo {author} {\bibfnamefont {C.}~\bibnamefont {Lange}}, \bibinfo
  {author} {\bibfnamefont {U.}~\bibnamefont {Huttner}}, \bibinfo {author}
  {\bibfnamefont {D.}~\bibnamefont {Golde}}, \bibinfo {author} {\bibfnamefont
  {T.}~\bibnamefont {Meier}}, \bibinfo {author} {\bibfnamefont
  {M.}~\bibnamefont {Kira}}, \bibinfo {author} {\bibfnamefont {S.~W.}\
  \bibnamefont {Koch}}, \ and\ \bibinfo {author} {\bibfnamefont
  {R.}~\bibnamefont {Huber}},\ }\bibfield  {title} {\enquote {\bibinfo {title}
  {Sub-cycle control of terahertz high-harmonic generation by dynamical bloch
  oscillations},}\ }\href {http://dx.doi.org/10.1038/nphoton.2013.349}
  {\bibfield  {journal} {\bibinfo  {journal} {Nat. Photonics}\ }\textbf
  {\bibinfo {volume} {8}},\ \bibinfo {pages} {119} (\bibinfo {year}
  {2014})}\BibitemShut {NoStop}%
\bibitem [{\citenamefont {Higuchi}\ \emph {et~al.}(2014)\citenamefont
  {Higuchi}, \citenamefont {Stockman},\ and\ \citenamefont
  {Hommelhoff}}]{Higuchi2014}%
  \BibitemOpen
  \bibfield  {author} {\bibinfo {author} {\bibfnamefont {T.}~\bibnamefont
  {Higuchi}}, \bibinfo {author} {\bibfnamefont {M.~I.}\ \bibnamefont
  {Stockman}}, \ and\ \bibinfo {author} {\bibfnamefont {P.}~\bibnamefont
  {Hommelhoff}},\ }\bibfield  {title} {\enquote {\bibinfo {title} {Strong-field
  perspective on high-harmonic radiation from bulk solids},}\ }\href {\doibase
  10.1103/PhysRevLett.113.213901} {\bibfield  {journal} {\bibinfo  {journal}
  {Phys. Rev. Lett.}\ }\textbf {\bibinfo {volume} {113}},\ \bibinfo {pages}
  {213901} (\bibinfo {year} {2014})}\BibitemShut {NoStop}%
\bibitem [{\citenamefont {Luu}\ \emph {et~al.}(2015)\citenamefont {Luu},
  \citenamefont {Garg}, \citenamefont {Kruchinin}, \citenamefont {Moulet},
  \citenamefont {Hassan},\ and\ \citenamefont {Goulielmakis}}]{Luu2015}%
  \BibitemOpen
  \bibfield  {author} {\bibinfo {author} {\bibfnamefont {T.~T.}\ \bibnamefont
  {Luu}}, \bibinfo {author} {\bibfnamefont {M.}~\bibnamefont {Garg}}, \bibinfo
  {author} {\bibfnamefont {S.~Y.}\ \bibnamefont {Kruchinin}}, \bibinfo {author}
  {\bibfnamefont {A.}~\bibnamefont {Moulet}}, \bibinfo {author} {\bibfnamefont
  {M.~T.}\ \bibnamefont {Hassan}}, \ and\ \bibinfo {author} {\bibfnamefont
  {E.}~\bibnamefont {Goulielmakis}},\ }\bibfield  {title} {\enquote {\bibinfo
  {title} {Extreme ultraviolet high-harmonic spectroscopy of solids},}\ }\href
  {https://doi.org/10.1038/nature14456} {\bibfield  {journal} {\bibinfo
  {journal} {Nature}\ }\textbf {\bibinfo {volume} {521}},\ \bibinfo {pages}
  {498 EP --} (\bibinfo {year} {2015})}\BibitemShut {NoStop}%
\bibitem [{\citenamefont {Wu}\ \emph {et~al.}(2015)\citenamefont {Wu},
  \citenamefont {Ghimire}, \citenamefont {Reis}, \citenamefont {Schafer},\ and\
  \citenamefont {Gaarde}}]{Wu2015}%
  \BibitemOpen
  \bibfield  {author} {\bibinfo {author} {\bibfnamefont {M.}~\bibnamefont
  {Wu}}, \bibinfo {author} {\bibfnamefont {S.}~\bibnamefont {Ghimire}},
  \bibinfo {author} {\bibfnamefont {D.~A.}\ \bibnamefont {Reis}}, \bibinfo
  {author} {\bibfnamefont {K.~J.}\ \bibnamefont {Schafer}}, \ and\ \bibinfo
  {author} {\bibfnamefont {M.~B.}\ \bibnamefont {Gaarde}},\ }\bibfield  {title}
  {\enquote {\bibinfo {title} {High-harmonic generation from bloch electrons in
  solids},}\ }\href {\doibase 10.1103/PhysRevA.91.043839} {\bibfield  {journal}
  {\bibinfo  {journal} {Phys. Rev. A}\ }\textbf {\bibinfo {volume} {91}},\
  \bibinfo {pages} {043839} (\bibinfo {year} {2015})}\BibitemShut {NoStop}%
\bibitem [{\citenamefont {Hohenleutner}\ \emph {et~al.}(2015)\citenamefont
  {Hohenleutner}, \citenamefont {Langer}, \citenamefont {Schubert},
  \citenamefont {Knorr}, \citenamefont {Huttner}, \citenamefont {Koch},
  \citenamefont {Kira},\ and\ \citenamefont {Huber}}]{Hohenleutner2015}%
  \BibitemOpen
  \bibfield  {author} {\bibinfo {author} {\bibfnamefont {M.}~\bibnamefont
  {Hohenleutner}}, \bibinfo {author} {\bibfnamefont {F.}~\bibnamefont
  {Langer}}, \bibinfo {author} {\bibfnamefont {O.}~\bibnamefont {Schubert}},
  \bibinfo {author} {\bibfnamefont {M.}~\bibnamefont {Knorr}}, \bibinfo
  {author} {\bibfnamefont {U.}~\bibnamefont {Huttner}}, \bibinfo {author}
  {\bibfnamefont {S.~W.}\ \bibnamefont {Koch}}, \bibinfo {author}
  {\bibfnamefont {M.}~\bibnamefont {Kira}}, \ and\ \bibinfo {author}
  {\bibfnamefont {R.}~\bibnamefont {Huber}},\ }\bibfield  {title} {\enquote
  {\bibinfo {title} {Real-time observation of interfering crystal electrons in
  high-harmonic generation},}\ }\href {https://doi.org/10.1038/nature14652}
  {\bibfield  {journal} {\bibinfo  {journal} {Nature}\ }\textbf {\bibinfo
  {volume} {523}},\ \bibinfo {pages} {572 EP --} (\bibinfo {year}
  {2015})}\BibitemShut {NoStop}%
\bibitem [{\citenamefont {Tamaya}\ \emph {et~al.}(2016)\citenamefont {Tamaya},
  \citenamefont {Ishikawa}, \citenamefont {Ogawa},\ and\ \citenamefont
  {Tanaka}}]{Tamaya2016}%
  \BibitemOpen
  \bibfield  {author} {\bibinfo {author} {\bibfnamefont {T.}~\bibnamefont
  {Tamaya}}, \bibinfo {author} {\bibfnamefont {A.}~\bibnamefont {Ishikawa}},
  \bibinfo {author} {\bibfnamefont {T.}~\bibnamefont {Ogawa}}, \ and\ \bibinfo
  {author} {\bibfnamefont {K.}~\bibnamefont {Tanaka}},\ }\bibfield  {title}
  {\enquote {\bibinfo {title} {Diabatic mechanisms of higher-order harmonic
  generation in solid-state materials under high-intensity electric fields},}\
  }\href {\doibase 10.1103/PhysRevLett.116.016601} {\bibfield  {journal}
  {\bibinfo  {journal} {Phys. Rev. Lett.}\ }\textbf {\bibinfo {volume} {116}},\
  \bibinfo {pages} {016601} (\bibinfo {year} {2016})}\BibitemShut {NoStop}%
\bibitem [{\citenamefont {Garg}\ \emph {et~al.}(2018)\citenamefont {Garg},
  \citenamefont {Kim},\ and\ \citenamefont {Goulielmakis}}]{Garg2018}%
  \BibitemOpen
  \bibfield  {author} {\bibinfo {author} {\bibfnamefont {M.}~\bibnamefont
  {Garg}}, \bibinfo {author} {\bibfnamefont {H.~Y.}\ \bibnamefont {Kim}}, \
  and\ \bibinfo {author} {\bibfnamefont {E.}~\bibnamefont {Goulielmakis}},\
  }\bibfield  {title} {\enquote {\bibinfo {title} {Ultimate waveform
  reproducibility of extreme-ultraviolet pulses by high-harmonic generation in
  quartz},}\ }\href {\doibase 10.1038/s41566-018-0123-6} {\bibfield  {journal}
  {\bibinfo  {journal} {Nat. Photonics}\ }\textbf {\bibinfo {volume} {12}},\
  \bibinfo {pages} {291--296} (\bibinfo {year} {2018})}\BibitemShut {NoStop}%
\bibitem [{\citenamefont {Li}\ \emph {et~al.}(2019{\natexlab{a}})\citenamefont
  {Li}, \citenamefont {Lan}, \citenamefont {Zhu}, \citenamefont {Huang},
  \citenamefont {Zhang}, \citenamefont {Lein},\ and\ \citenamefont
  {Lu}}]{Li2019}%
  \BibitemOpen
  \bibfield  {author} {\bibinfo {author} {\bibfnamefont {L.}~\bibnamefont
  {Li}}, \bibinfo {author} {\bibfnamefont {P.}~\bibnamefont {Lan}}, \bibinfo
  {author} {\bibfnamefont {X.}~\bibnamefont {Zhu}}, \bibinfo {author}
  {\bibfnamefont {T.}~\bibnamefont {Huang}}, \bibinfo {author} {\bibfnamefont
  {Q.}~\bibnamefont {Zhang}}, \bibinfo {author} {\bibfnamefont
  {M.}~\bibnamefont {Lein}}, \ and\ \bibinfo {author} {\bibfnamefont
  {P.}~\bibnamefont {Lu}},\ }\bibfield  {title} {\enquote {\bibinfo {title}
  {Reciprocal-space-trajectory perspective on high-harmonic generation in
  solids},}\ }\href {\doibase 10.1103/PhysRevLett.122.193901} {\bibfield
  {journal} {\bibinfo  {journal} {Phys. Rev. Lett.}\ }\textbf {\bibinfo
  {volume} {122}},\ \bibinfo {pages} {193901} (\bibinfo {year}
  {2019}{\natexlab{a}})}\BibitemShut {NoStop}%
\bibitem [{\citenamefont {Navarrete}\ \emph {et~al.}(2019)\citenamefont
  {Navarrete}, \citenamefont {Ciappina},\ and\ \citenamefont
  {Thumm}}]{Navarette2019}%
  \BibitemOpen
  \bibfield  {author} {\bibinfo {author} {\bibfnamefont {F.}~\bibnamefont
  {Navarrete}}, \bibinfo {author} {\bibfnamefont {M.~F.}\ \bibnamefont
  {Ciappina}}, \ and\ \bibinfo {author} {\bibfnamefont {U.}~\bibnamefont
  {Thumm}},\ }\bibfield  {title} {\enquote {\bibinfo {title}
  {Crystal-momentum-resolved contributions to high-order harmonic generation in
  solids},}\ }\href {\doibase 10.1103/PhysRevA.100.033405} {\bibfield
  {journal} {\bibinfo  {journal} {Phys. Rev. A}\ }\textbf {\bibinfo {volume}
  {100}},\ \bibinfo {pages} {033405} (\bibinfo {year} {2019})}\BibitemShut
  {NoStop}%
\bibitem [{\citenamefont {Vampa}\ \emph
  {et~al.}(2015{\natexlab{c}})\citenamefont {Vampa}, \citenamefont {McDonald},
  \citenamefont {Orlando}, \citenamefont {Corkum},\ and\ \citenamefont
  {Brabec}}]{Vampa2015}%
  \BibitemOpen
  \bibfield  {author} {\bibinfo {author} {\bibfnamefont {G.}~\bibnamefont
  {Vampa}}, \bibinfo {author} {\bibfnamefont {C.~R.}\ \bibnamefont {McDonald}},
  \bibinfo {author} {\bibfnamefont {G.}~\bibnamefont {Orlando}}, \bibinfo
  {author} {\bibfnamefont {P.~B.}\ \bibnamefont {Corkum}}, \ and\ \bibinfo
  {author} {\bibfnamefont {T.}~\bibnamefont {Brabec}},\ }\bibfield  {title}
  {\enquote {\bibinfo {title} {Semiclassical analysis of high harmonic
  generation in bulk crystals},}\ }\href {\doibase 10.1103/PhysRevB.91.064302}
  {\bibfield  {journal} {\bibinfo  {journal} {Phys. Rev. B}\ }\textbf {\bibinfo
  {volume} {91}},\ \bibinfo {pages} {064302} (\bibinfo {year}
  {2015}{\natexlab{c}})}\BibitemShut {NoStop}%
\bibitem [{\citenamefont {Yu}\ \emph {et~al.}(2019)\citenamefont {Yu},
  \citenamefont {Hansen},\ and\ \citenamefont {Madsen}}]{Yu2019}%
  \BibitemOpen
  \bibfield  {author} {\bibinfo {author} {\bibfnamefont {C.}~\bibnamefont
  {Yu}}, \bibinfo {author} {\bibfnamefont {K.~K.}\ \bibnamefont {Hansen}}, \
  and\ \bibinfo {author} {\bibfnamefont {L.~B.}\ \bibnamefont {Madsen}},\
  }\bibfield  {title} {\enquote {\bibinfo {title} {Enhanced high-order harmonic
  generation in donor-doped band-gap materials},}\ }\href {\doibase
  10.1103/PhysRevA.99.013435} {\bibfield  {journal} {\bibinfo  {journal} {Phys.
  Rev. A}\ }\textbf {\bibinfo {volume} {99}},\ \bibinfo {pages} {013435}
  (\bibinfo {year} {2019})}\BibitemShut {NoStop}%
\bibitem [{\citenamefont {Ikemachi}\ \emph {et~al.}(2017)\citenamefont
  {Ikemachi}, \citenamefont {Shinohara}, \citenamefont {Sato}, \citenamefont
  {Yumoto}, \citenamefont {Kuwata-Gonokami},\ and\ \citenamefont
  {Ishikawa}}]{Ikemachi2017}%
  \BibitemOpen
  \bibfield  {author} {\bibinfo {author} {\bibfnamefont {T.}~\bibnamefont
  {Ikemachi}}, \bibinfo {author} {\bibfnamefont {Y.}~\bibnamefont {Shinohara}},
  \bibinfo {author} {\bibfnamefont {T.}~\bibnamefont {Sato}}, \bibinfo {author}
  {\bibfnamefont {J.}~\bibnamefont {Yumoto}}, \bibinfo {author} {\bibfnamefont
  {M.}~\bibnamefont {Kuwata-Gonokami}}, \ and\ \bibinfo {author} {\bibfnamefont
  {K.~L.}\ \bibnamefont {Ishikawa}},\ }\bibfield  {title} {\enquote {\bibinfo
  {title} {Trajectory analysis of high-order-harmonic generation from periodic
  crystals},}\ }\href {\doibase 10.1103/PhysRevA.95.043416} {\bibfield
  {journal} {\bibinfo  {journal} {Phys. Rev. A}\ }\textbf {\bibinfo {volume}
  {95}},\ \bibinfo {pages} {043416} (\bibinfo {year} {2017})}\BibitemShut
  {NoStop}%
\bibitem [{\citenamefont {Tancogne-Dejean}\ \emph {et~al.}(2017)\citenamefont
  {Tancogne-Dejean}, \citenamefont {M{\"u}cke}, \citenamefont {K{\"a}rtner},\
  and\ \citenamefont {Rubio}}]{Tancogne-Dejean2017}%
  \BibitemOpen
  \bibfield  {author} {\bibinfo {author} {\bibfnamefont {N.}~\bibnamefont
  {Tancogne-Dejean}}, \bibinfo {author} {\bibfnamefont {O.~D.}\ \bibnamefont
  {M{\"u}cke}}, \bibinfo {author} {\bibfnamefont {F.~X.}\ \bibnamefont
  {K{\"a}rtner}}, \ and\ \bibinfo {author} {\bibfnamefont {A.}~\bibnamefont
  {Rubio}},\ }\bibfield  {title} {\enquote {\bibinfo {title} {Ellipticity
  dependence of high-harmonic generation in solids originating from coupled
  intraband and interband dynamics},}\ }\href {\doibase
  10.1038/s41467-017-00764-5} {\bibfield  {journal} {\bibinfo  {journal} {Nat.
  Commun.}\ }\textbf {\bibinfo {volume} {8}},\ \bibinfo {pages} {745} (\bibinfo
  {year} {2017})}\BibitemShut {NoStop}%
\bibitem [{\citenamefont {Tancogne-Dejean}\ and\ \citenamefont
  {Rubio}(2018)}]{Tancogne-Dejean2018}%
  \BibitemOpen
  \bibfield  {author} {\bibinfo {author} {\bibfnamefont {N.}~\bibnamefont
  {Tancogne-Dejean}}\ and\ \bibinfo {author} {\bibfnamefont {A.}~\bibnamefont
  {Rubio}},\ }\bibfield  {title} {\enquote {\bibinfo {title} {Atomic-like
  high-harmonic generation from two-dimensional materials},}\ }\href {\doibase
  10.1126/sciadv.aao5207} {\bibfield  {journal} {\bibinfo  {journal} {Sci.
  Adv.}\ }\textbf {\bibinfo {volume} {4}} (\bibinfo {year} {2018}),\
  10.1126/sciadv.aao5207}\BibitemShut {NoStop}%
\bibitem [{\citenamefont {Yoshikawa}\ \emph {et~al.}(2019)\citenamefont
  {Yoshikawa}, \citenamefont {Nagai}, \citenamefont {Uchida}, \citenamefont
  {Takaguchi}, \citenamefont {Sasaki}, \citenamefont {Miyata},\ and\
  \citenamefont {Tanaka}}]{Yoshikawa2019}%
  \BibitemOpen
  \bibfield  {author} {\bibinfo {author} {\bibfnamefont {N.}~\bibnamefont
  {Yoshikawa}}, \bibinfo {author} {\bibfnamefont {K.}~\bibnamefont {Nagai}},
  \bibinfo {author} {\bibfnamefont {K.}~\bibnamefont {Uchida}}, \bibinfo
  {author} {\bibfnamefont {Y.}~\bibnamefont {Takaguchi}}, \bibinfo {author}
  {\bibfnamefont {S.}~\bibnamefont {Sasaki}}, \bibinfo {author} {\bibfnamefont
  {Y.}~\bibnamefont {Miyata}}, \ and\ \bibinfo {author} {\bibfnamefont
  {K.}~\bibnamefont {Tanaka}},\ }\bibfield  {title} {\enquote {\bibinfo {title}
  {Interband resonant high-harmonic generation by valley polarized
  electron-hole pairs},}\ }\href {\doibase 10.1038/s41467-019-11697-6}
  {\bibfield  {journal} {\bibinfo  {journal} {Nat. Commun.}\ }\textbf {\bibinfo
  {volume} {10}},\ \bibinfo {pages} {3709} (\bibinfo {year}
  {2019})}\BibitemShut {NoStop}%
\bibitem [{\citenamefont {Corkum}(1993)}]{Corkum1993}%
  \BibitemOpen
  \bibfield  {author} {\bibinfo {author} {\bibfnamefont {P.~B.}\ \bibnamefont
  {Corkum}},\ }\bibfield  {title} {\enquote {\bibinfo {title} {Plasma
  perspective on strong field multiphoton ionization},}\ }\href {\doibase
  10.1103/PhysRevLett.71.1994} {\bibfield  {journal} {\bibinfo  {journal}
  {Phys. Rev. Lett.}\ }\textbf {\bibinfo {volume} {71}},\ \bibinfo {pages}
  {1994--1997} (\bibinfo {year} {1993})}\BibitemShut {NoStop}%
\bibitem [{\citenamefont {Lewenstein}\ \emph {et~al.}(1994)\citenamefont
  {Lewenstein}, \citenamefont {Balcou}, \citenamefont {Ivanov}, \citenamefont
  {L'Huillier},\ and\ \citenamefont {Corkum}}]{Lewenstein1994}%
  \BibitemOpen
  \bibfield  {author} {\bibinfo {author} {\bibfnamefont {M.}~\bibnamefont
  {Lewenstein}}, \bibinfo {author} {\bibfnamefont {P.}~\bibnamefont {Balcou}},
  \bibinfo {author} {\bibfnamefont {M.~Y.}\ \bibnamefont {Ivanov}}, \bibinfo
  {author} {\bibfnamefont {A.}~\bibnamefont {L'Huillier}}, \ and\ \bibinfo
  {author} {\bibfnamefont {P.~B.}\ \bibnamefont {Corkum}},\ }\bibfield  {title}
  {\enquote {\bibinfo {title} {Theory of high-harmonic generation by
  low-frequency laser fields},}\ }\href {\doibase 10.1103/PhysRevA.49.2117}
  {\bibfield  {journal} {\bibinfo  {journal} {Phys. Rev. A}\ }\textbf {\bibinfo
  {volume} {49}},\ \bibinfo {pages} {2117} (\bibinfo {year}
  {1994})}\BibitemShut {NoStop}%
\bibitem [{\citenamefont {Jiang}\ \emph {et~al.}(2018)\citenamefont {Jiang},
  \citenamefont {Chen}, \citenamefont {Wei}, \citenamefont {Yu}, \citenamefont
  {Lu},\ and\ \citenamefont {Lin}}]{Jiang2018}%
  \BibitemOpen
  \bibfield  {author} {\bibinfo {author} {\bibfnamefont {S.}~\bibnamefont
  {Jiang}}, \bibinfo {author} {\bibfnamefont {J.}~\bibnamefont {Chen}},
  \bibinfo {author} {\bibfnamefont {H.}~\bibnamefont {Wei}}, \bibinfo {author}
  {\bibfnamefont {C.}~\bibnamefont {Yu}}, \bibinfo {author} {\bibfnamefont
  {R.}~\bibnamefont {Lu}}, \ and\ \bibinfo {author} {\bibfnamefont {C.~D.}\
  \bibnamefont {Lin}},\ }\bibfield  {title} {\enquote {\bibinfo {title} {Role
  of the transition dipole amplitude and phase on the generation of odd and
  even high-order harmonics in crystals},}\ }\href {\doibase
  10.1103/PhysRevLett.120.253201} {\bibfield  {journal} {\bibinfo  {journal}
  {Phys. Rev. Lett.}\ }\textbf {\bibinfo {volume} {120}},\ \bibinfo {pages}
  {253201} (\bibinfo {year} {2018})}\BibitemShut {NoStop}%
\bibitem [{\citenamefont {Golde}\ \emph {et~al.}(2008)\citenamefont {Golde},
  \citenamefont {Meier},\ and\ \citenamefont {Koch}}]{Golde2008}%
  \BibitemOpen
  \bibfield  {author} {\bibinfo {author} {\bibfnamefont {D.}~\bibnamefont
  {Golde}}, \bibinfo {author} {\bibfnamefont {T.}~\bibnamefont {Meier}}, \ and\
  \bibinfo {author} {\bibfnamefont {S.~W.}\ \bibnamefont {Koch}},\ }\bibfield
  {title} {\enquote {\bibinfo {title} {High harmonics generated in
  semiconductor nanostructures by the coupled dynamics of optical inter- and
  intraband excitations},}\ }\href {\doibase 10.1103/PhysRevB.77.075330}
  {\bibfield  {journal} {\bibinfo  {journal} {Phys. Rev. B}\ }\textbf {\bibinfo
  {volume} {77}},\ \bibinfo {pages} {075330} (\bibinfo {year}
  {2008})}\BibitemShut {NoStop}%
\bibitem [{\citenamefont {Kira}\ and\ \citenamefont {Koch}(2012)}]{Kira2012}%
  \BibitemOpen
  \bibfield  {author} {\bibinfo {author} {\bibfnamefont {M.}~\bibnamefont
  {Kira}}\ and\ \bibinfo {author} {\bibfnamefont {S.~W.}\ \bibnamefont
  {Koch}},\ }\href@noop {} {\emph {\bibinfo {title} {Semiconductor Quantum
  Optics}}}\ (\bibinfo  {publisher} {Cambridge University Press},\ \bibinfo
  {year} {2012})\BibitemShut {NoStop}%
\bibitem [{\citenamefont {Yoshikawa}\ \emph {et~al.}(2017)\citenamefont
  {Yoshikawa}, \citenamefont {Tamaya},\ and\ \citenamefont
  {Tanaka}}]{Yoshikawa2017}%
  \BibitemOpen
  \bibfield  {author} {\bibinfo {author} {\bibfnamefont {N.}~\bibnamefont
  {Yoshikawa}}, \bibinfo {author} {\bibfnamefont {T.}~\bibnamefont {Tamaya}}, \
  and\ \bibinfo {author} {\bibfnamefont {K.}~\bibnamefont {Tanaka}},\
  }\bibfield  {title} {\enquote {\bibinfo {title} {High-harmonic generation in
  graphene enhanced by elliptically polarized light excitation},}\ }\href
  {\doibase 10.1126/science.aam8861} {\bibfield  {journal} {\bibinfo  {journal}
  {Science}\ }\textbf {\bibinfo {volume} {356}},\ \bibinfo {pages} {736--738}
  (\bibinfo {year} {2017})}\BibitemShut {NoStop}%
\bibitem [{\citenamefont {Taucer}\ \emph {et~al.}(2017)\citenamefont {Taucer},
  \citenamefont {Hammond}, \citenamefont {Corkum}, \citenamefont {Vampa},
  \citenamefont {Couture}, \citenamefont {Thir\'e}, \citenamefont {Schmidt},
  \citenamefont {L\'egar\'e}, \citenamefont {Selvi}, \citenamefont {Unsuree},
  \citenamefont {Hamilton}, \citenamefont {Echtermeyer},\ and\ \citenamefont
  {Denecke}}]{Taucer2017}%
  \BibitemOpen
  \bibfield  {author} {\bibinfo {author} {\bibfnamefont {M.}~\bibnamefont
  {Taucer}}, \bibinfo {author} {\bibfnamefont {T.~J.}\ \bibnamefont {Hammond}},
  \bibinfo {author} {\bibfnamefont {P.~B.}\ \bibnamefont {Corkum}}, \bibinfo
  {author} {\bibfnamefont {G.}~\bibnamefont {Vampa}}, \bibinfo {author}
  {\bibfnamefont {C.}~\bibnamefont {Couture}}, \bibinfo {author} {\bibfnamefont
  {N.}~\bibnamefont {Thir\'e}}, \bibinfo {author} {\bibfnamefont {B.~E.}\
  \bibnamefont {Schmidt}}, \bibinfo {author} {\bibfnamefont {F.}~\bibnamefont
  {L\'egar\'e}}, \bibinfo {author} {\bibfnamefont {H.}~\bibnamefont {Selvi}},
  \bibinfo {author} {\bibfnamefont {N.}~\bibnamefont {Unsuree}}, \bibinfo
  {author} {\bibfnamefont {B.}~\bibnamefont {Hamilton}}, \bibinfo {author}
  {\bibfnamefont {T.~J.}\ \bibnamefont {Echtermeyer}}, \ and\ \bibinfo {author}
  {\bibfnamefont {M.~A.}\ \bibnamefont {Denecke}},\ }\bibfield  {title}
  {\enquote {\bibinfo {title} {Nonperturbative harmonic generation in graphene
  from intense midinfrared pulsed light},}\ }\href {\doibase
  10.1103/PhysRevB.96.195420} {\bibfield  {journal} {\bibinfo  {journal} {Phys.
  Rev. B}\ }\textbf {\bibinfo {volume} {96}},\ \bibinfo {pages} {195420}
  (\bibinfo {year} {2017})}\BibitemShut {NoStop}%
\bibitem [{\citenamefont {Hafez}\ \emph {et~al.}(2018)\citenamefont {Hafez},
  \citenamefont {Kovalev}, \citenamefont {Deinert}, \citenamefont {Mics},
  \citenamefont {Green}, \citenamefont {Awari}, \citenamefont {Chen},
  \citenamefont {Germanskiy}, \citenamefont {Lehnert}, \citenamefont
  {Teichert}, \citenamefont {Wang}, \citenamefont {Tielrooij}, \citenamefont
  {Liu}, \citenamefont {Chen}, \citenamefont {Narita}, \citenamefont
  {M{\"u}llen}, \citenamefont {Bonn}, \citenamefont {Gensch},\ and\
  \citenamefont {Turchinovich}}]{Hafez2018}%
  \BibitemOpen
  \bibfield  {author} {\bibinfo {author} {\bibfnamefont {H.~A.}\ \bibnamefont
  {Hafez}}, \bibinfo {author} {\bibfnamefont {S.}~\bibnamefont {Kovalev}},
  \bibinfo {author} {\bibfnamefont {J.-C.}\ \bibnamefont {Deinert}}, \bibinfo
  {author} {\bibfnamefont {Z.}~\bibnamefont {Mics}}, \bibinfo {author}
  {\bibfnamefont {B.}~\bibnamefont {Green}}, \bibinfo {author} {\bibfnamefont
  {N.}~\bibnamefont {Awari}}, \bibinfo {author} {\bibfnamefont
  {M.}~\bibnamefont {Chen}}, \bibinfo {author} {\bibfnamefont {S.}~\bibnamefont
  {Germanskiy}}, \bibinfo {author} {\bibfnamefont {U.}~\bibnamefont {Lehnert}},
  \bibinfo {author} {\bibfnamefont {J.}~\bibnamefont {Teichert}}, \bibinfo
  {author} {\bibfnamefont {Z.}~\bibnamefont {Wang}}, \bibinfo {author}
  {\bibfnamefont {K.-J.}\ \bibnamefont {Tielrooij}}, \bibinfo {author}
  {\bibfnamefont {Z.}~\bibnamefont {Liu}}, \bibinfo {author} {\bibfnamefont
  {Z.}~\bibnamefont {Chen}}, \bibinfo {author} {\bibfnamefont {A.}~\bibnamefont
  {Narita}}, \bibinfo {author} {\bibfnamefont {K.}~\bibnamefont {M{\"u}llen}},
  \bibinfo {author} {\bibfnamefont {M.}~\bibnamefont {Bonn}}, \bibinfo {author}
  {\bibfnamefont {M.}~\bibnamefont {Gensch}}, \ and\ \bibinfo {author}
  {\bibfnamefont {D.}~\bibnamefont {Turchinovich}},\ }\bibfield  {title}
  {\enquote {\bibinfo {title} {Extremely efficient terahertz high-harmonic
  generation in graphene by hot dirac fermions},}\ }\href {\doibase
  10.1038/s41586-018-0508-1} {\bibfield  {journal} {\bibinfo  {journal}
  {Nature}\ }\textbf {\bibinfo {volume} {561}},\ \bibinfo {pages} {507--511}
  (\bibinfo {year} {2018})}\BibitemShut {NoStop}%
\bibitem [{\citenamefont {Silva}\ \emph
  {et~al.}(2019{\natexlab{b}})\citenamefont {Silva}, \citenamefont {Mart\'in},\
  and\ \citenamefont {Ivanov}}]{Silva2019b}%
  \BibitemOpen
  \bibfield  {author} {\bibinfo {author} {\bibfnamefont {R.}~\bibnamefont
  {Silva}}, \bibinfo {author} {\bibfnamefont {F.}~\bibnamefont {Mart\'in}}, \
  and\ \bibinfo {author} {\bibfnamefont {M.}~\bibnamefont {Ivanov}},\
  }\bibfield  {title} {\enquote {\bibinfo {title} {High harmonic generation in
  crystals using maximally localized wannier functions},}\ }\href@noop {}
  {\bibfield  {journal} {\bibinfo  {journal} {arXiv:1904.00283v2}\ } (\bibinfo
  {year} {2019}{\natexlab{b}})}\BibitemShut {NoStop}%
\bibitem [{\citenamefont {Mrudul}\ \emph {et~al.}(2019)\citenamefont {Mrudul},
  \citenamefont {Tancogne-Dejean}, \citenamefont {Rubio},\ and\ \citenamefont
  {Dixit}}]{Mrudul2019}%
  \BibitemOpen
  \bibfield  {author} {\bibinfo {author} {\bibfnamefont {M.~S.}\ \bibnamefont
  {Mrudul}}, \bibinfo {author} {\bibfnamefont {N.}~\bibnamefont
  {Tancogne-Dejean}}, \bibinfo {author} {\bibfnamefont {A.}~\bibnamefont
  {Rubio}}, \ and\ \bibinfo {author} {\bibfnamefont {G.}~\bibnamefont
  {Dixit}},\ }\bibfield  {title} {\enquote {\bibinfo {title} {High-harmonic
  generation from spin-polarised defects in solids},}\ }\href@noop {}
  {\bibfield  {journal} {\bibinfo  {journal} {arXiv:1906.10224}\ } (\bibinfo
  {year} {2019})}\BibitemShut {NoStop}%
\bibitem [{\citenamefont {Floss}\ \emph {et~al.}(2018)\citenamefont {Floss},
  \citenamefont {Lemell}, \citenamefont {Wachter}, \citenamefont {Smejkal},
  \citenamefont {Sato}, \citenamefont {Tong}, \citenamefont {Yabana},\ and\
  \citenamefont {Burgd\"orfer}}]{Floss2018}%
  \BibitemOpen
  \bibfield  {author} {\bibinfo {author} {\bibfnamefont {I.}~\bibnamefont
  {Floss}}, \bibinfo {author} {\bibfnamefont {C.}~\bibnamefont {Lemell}},
  \bibinfo {author} {\bibfnamefont {G.}~\bibnamefont {Wachter}}, \bibinfo
  {author} {\bibfnamefont {V.}~\bibnamefont {Smejkal}}, \bibinfo {author}
  {\bibfnamefont {S.~A.}\ \bibnamefont {Sato}}, \bibinfo {author}
  {\bibfnamefont {X.-M.}\ \bibnamefont {Tong}}, \bibinfo {author}
  {\bibfnamefont {K.}~\bibnamefont {Yabana}}, \ and\ \bibinfo {author}
  {\bibfnamefont {J.}~\bibnamefont {Burgd\"orfer}},\ }\bibfield  {title}
  {\enquote {\bibinfo {title} {Ab initio multiscale simulation of high-order
  harmonic generation in solids},}\ }\href {\doibase
  10.1103/PhysRevA.97.011401} {\bibfield  {journal} {\bibinfo  {journal} {Phys.
  Rev. A}\ }\textbf {\bibinfo {volume} {97}},\ \bibinfo {pages} {011401(R)}
  (\bibinfo {year} {2018})}\BibitemShut {NoStop}%
\bibitem [{\citenamefont {Lu}\ \emph {et~al.}(2019)\citenamefont {Lu},
  \citenamefont {Cunningham}, \citenamefont {You}, \citenamefont {Reis},\ and\
  \citenamefont {Ghimire}}]{Lu2019}%
  \BibitemOpen
  \bibfield  {author} {\bibinfo {author} {\bibfnamefont {J.}~\bibnamefont
  {Lu}}, \bibinfo {author} {\bibfnamefont {E.~F.}\ \bibnamefont {Cunningham}},
  \bibinfo {author} {\bibfnamefont {Y.~S.}\ \bibnamefont {You}}, \bibinfo
  {author} {\bibfnamefont {D.~A.}\ \bibnamefont {Reis}}, \ and\ \bibinfo
  {author} {\bibfnamefont {S.}~\bibnamefont {Ghimire}},\ }\bibfield  {title}
  {\enquote {\bibinfo {title} {Interferometry of dipole phase in high harmonics
  from solids},}\ }\href {\doibase 10.1038/s41566-018-0326-x} {\bibfield
  {journal} {\bibinfo  {journal} {Nat. Photonics}\ }\textbf {\bibinfo {volume}
  {13}},\ \bibinfo {pages} {96} (\bibinfo {year} {2019})}\BibitemShut {NoStop}%
\bibitem [{\citenamefont {Taghizadeh}\ \emph {et~al.}(2017)\citenamefont
  {Taghizadeh}, \citenamefont {Hipolito},\ and\ \citenamefont
  {Pedersen}}]{Taghizadeh2017}%
  \BibitemOpen
  \bibfield  {author} {\bibinfo {author} {\bibfnamefont {A.}~\bibnamefont
  {Taghizadeh}}, \bibinfo {author} {\bibfnamefont {F.}~\bibnamefont
  {Hipolito}}, \ and\ \bibinfo {author} {\bibfnamefont {T.~G.}\ \bibnamefont
  {Pedersen}},\ }\bibfield  {title} {\enquote {\bibinfo {title} {Linear and
  nonlinear optical response of crystals using length and velocity gauges:
  Effect of basis truncation},}\ }\href {\doibase 10.1103/PhysRevB.96.195413}
  {\bibfield  {journal} {\bibinfo  {journal} {Phys. Rev. B}\ }\textbf {\bibinfo
  {volume} {96}},\ \bibinfo {pages} {195413} (\bibinfo {year}
  {2017})}\BibitemShut {NoStop}%
\bibitem [{sup()}]{suppmat2020_imperf_recol}%
  \BibitemOpen
  \href@noop {} {}\bibinfo {note} {See Supplemental Material at [URL] for
  additional details about the structure calculations, the SBEs, the choice of
  $R_0$, and the method used in the calculation of the quantum wave packet
  spreading.}\BibitemShut {Stop}%
\bibitem [{\citenamefont {Vanderbilt}(2018)}]{Vanderbilt2018}%
  \BibitemOpen
  \bibfield  {author} {\bibinfo {author} {\bibfnamefont {D.}~\bibnamefont
  {Vanderbilt}},\ }\href@noop {} {\emph {\bibinfo {title} {Berry phases in
  electronic structure theory: electric polarization, orbital magnetization and
  topological insulators}}}\ (\bibinfo  {publisher} {Cambridge University
  Press},\ \bibinfo {year} {2018})\BibitemShut {NoStop}%
\bibitem [{Note1()}]{Note1}%
  \BibitemOpen
  \bibinfo {note} {Note that similar equations appear in \cite {Li2019phase},
  but without inclusion of the electron-hole polarization energy. They were
  applied to a model one-dimensional system where the Berry connections can
  naturally chosen to be zero. We stress that imperfect recollisions can only
  occur in systems with dimensions higher than one.}\BibitemShut {Stop}%
\bibitem [{Note2()}]{Note2}%
  \BibitemOpen
  \bibinfo {note} {This is consistent with the discussion of laser dressing in
  Ref.~\cite {Tamaya2016}.}\BibitemShut {Stop}%
\bibitem [{\citenamefont {Hipolito}\ \emph {et~al.}(2016)\citenamefont
  {Hipolito}, \citenamefont {Pedersen},\ and\ \citenamefont
  {Pereira}}]{Hipolito2016}%
  \BibitemOpen
  \bibfield  {author} {\bibinfo {author} {\bibfnamefont {F.}~\bibnamefont
  {Hipolito}}, \bibinfo {author} {\bibfnamefont {T.~G.}\ \bibnamefont
  {Pedersen}}, \ and\ \bibinfo {author} {\bibfnamefont {V.~M.}\ \bibnamefont
  {Pereira}},\ }\bibfield  {title} {\enquote {\bibinfo {title} {Nonlinear
  photocurrents in two-dimensional systems based on graphene and boron
  nitride},}\ }\href {\doibase 10.1103/PhysRevB.94.045434} {\bibfield
  {journal} {\bibinfo  {journal} {Phys. Rev. B}\ }\textbf {\bibinfo {volume}
  {94}},\ \bibinfo {pages} {045434} (\bibinfo {year} {2016})}\BibitemShut
  {NoStop}%
\bibitem [{\citenamefont {Van~Hove}(1953)}]{vanHove1953}%
  \BibitemOpen
  \bibfield  {author} {\bibinfo {author} {\bibfnamefont {L.}~\bibnamefont
  {Van~Hove}},\ }\bibfield  {title} {\enquote {\bibinfo {title} {The occurrence
  of singularities in the elastic frequency distribution of a crystal},}\
  }\href {\doibase 10.1103/PhysRev.89.1189} {\bibfield  {journal} {\bibinfo
  {journal} {Phys. Rev.}\ }\textbf {\bibinfo {volume} {89}},\ \bibinfo {pages}
  {1189} (\bibinfo {year} {1953})}\BibitemShut {NoStop}%
\bibitem [{\citenamefont {Uzan}\ \emph {et~al.}(2019)\citenamefont {Uzan},
  \citenamefont {Orenstein}, \citenamefont {Jim\'enez-Gal\'an}, \citenamefont
  {McDonald}, \citenamefont {Silva}, \citenamefont {Bruner}, \citenamefont
  {Klimkin}, \citenamefont {Blanchet}, \citenamefont {Arusi-Parpar},
  \citenamefont {Kr\"uger}, \citenamefont {Rubtsov}, \citenamefont {Smirnova},
  \citenamefont {Ivanov}, \citenamefont {Yan}, \citenamefont {Brabec},\ and\
  \citenamefont {Dudovich}}]{Uzan2018}%
  \BibitemOpen
  \bibfield  {author} {\bibinfo {author} {\bibfnamefont {A.~J.}\ \bibnamefont
  {Uzan}}, \bibinfo {author} {\bibfnamefont {G.}~\bibnamefont {Orenstein}},
  \bibinfo {author} {\bibfnamefont {A.}~\bibnamefont {Jim\'enez-Gal\'an}},
  \bibinfo {author} {\bibfnamefont {C.}~\bibnamefont {McDonald}}, \bibinfo
  {author} {\bibfnamefont {R.~E.~F.}\ \bibnamefont {Silva}}, \bibinfo {author}
  {\bibfnamefont {B.~D.}\ \bibnamefont {Bruner}}, \bibinfo {author}
  {\bibfnamefont {N.~D.}\ \bibnamefont {Klimkin}}, \bibinfo {author}
  {\bibfnamefont {V.}~\bibnamefont {Blanchet}}, \bibinfo {author}
  {\bibfnamefont {T.}~\bibnamefont {Arusi-Parpar}}, \bibinfo {author}
  {\bibfnamefont {M.}~\bibnamefont {Kr\"uger}}, \bibinfo {author}
  {\bibfnamefont {A.~N.}\ \bibnamefont {Rubtsov}}, \bibinfo {author}
  {\bibfnamefont {O.}~\bibnamefont {Smirnova}}, \bibinfo {author}
  {\bibfnamefont {M.}~\bibnamefont {Ivanov}}, \bibinfo {author} {\bibfnamefont
  {B.}~\bibnamefont {Yan}}, \bibinfo {author} {\bibfnamefont {T.}~\bibnamefont
  {Brabec}}, \ and\ \bibinfo {author} {\bibfnamefont {N.}~\bibnamefont
  {Dudovich}},\ }\bibfield  {title} {\enquote {\bibinfo {title} {Multi-band
  petahertz currents resolved via high harmonic generation spectroscopy},}\
  }\href@noop {} {\bibfield  {journal} {\bibinfo  {journal}
  {arXiv:1904.00283v2}\ } (\bibinfo {year} {2019})}\BibitemShut {NoStop}%
\bibitem [{\citenamefont {Houston}(1940)}]{Houston1940}%
  \BibitemOpen
  \bibfield  {author} {\bibinfo {author} {\bibfnamefont {W.~V.}\ \bibnamefont
  {Houston}},\ }\bibfield  {title} {\enquote {\bibinfo {title} {Acceleration of
  electrons in a crystal lattice},}\ }\href {\doibase 10.1103/PhysRev.57.184}
  {\bibfield  {journal} {\bibinfo  {journal} {Phys. Rev.}\ }\textbf {\bibinfo
  {volume} {57}},\ \bibinfo {pages} {184--186} (\bibinfo {year}
  {1940})}\BibitemShut {NoStop}%
\bibitem [{\citenamefont {Krieger}\ and\ \citenamefont
  {Iafrate}(1986)}]{Krieger1986}%
  \BibitemOpen
  \bibfield  {author} {\bibinfo {author} {\bibfnamefont {J.~B.}\ \bibnamefont
  {Krieger}}\ and\ \bibinfo {author} {\bibfnamefont {G.~J.}\ \bibnamefont
  {Iafrate}},\ }\bibfield  {title} {\enquote {\bibinfo {title} {Time evolution
  of bloch electrons in a homogeneous electric field},}\ }\href {\doibase
  10.1103/PhysRevB.33.5494} {\bibfield  {journal} {\bibinfo  {journal} {Phys.
  Rev. B}\ }\textbf {\bibinfo {volume} {33}},\ \bibinfo {pages} {5494}
  (\bibinfo {year} {1986})}\BibitemShut {NoStop}%
\bibitem [{\citenamefont {Abadie}\ \emph {et~al.}(2018)\citenamefont {Abadie},
  \citenamefont {Wu},\ and\ \citenamefont {Gaarde}}]{Abadie2018}%
  \BibitemOpen
  \bibfield  {author} {\bibinfo {author} {\bibfnamefont {C.~Q.}\ \bibnamefont
  {Abadie}}, \bibinfo {author} {\bibfnamefont {M.}~\bibnamefont {Wu}}, \ and\
  \bibinfo {author} {\bibfnamefont {M.~B.}\ \bibnamefont {Gaarde}},\ }\bibfield
   {title} {\enquote {\bibinfo {title} {Spatiotemporal filtering of high
  harmonics in solids},}\ }\href {\doibase 10.1364/OL.43.005339} {\bibfield
  {journal} {\bibinfo  {journal} {Opt. Lett.}\ }\textbf {\bibinfo {volume}
  {43}},\ \bibinfo {pages} {5339} (\bibinfo {year} {2018})}\BibitemShut
  {NoStop}%
\bibitem [{\citenamefont {Huang}\ \emph {et~al.}(2017)\citenamefont {Huang},
  \citenamefont {Zhu}, \citenamefont {Li}, \citenamefont {Liu}, \citenamefont
  {Lan},\ and\ \citenamefont {Lu}}]{Huang2017}%
  \BibitemOpen
  \bibfield  {author} {\bibinfo {author} {\bibfnamefont {T.}~\bibnamefont
  {Huang}}, \bibinfo {author} {\bibfnamefont {X.}~\bibnamefont {Zhu}}, \bibinfo
  {author} {\bibfnamefont {L.}~\bibnamefont {Li}}, \bibinfo {author}
  {\bibfnamefont {X.}~\bibnamefont {Liu}}, \bibinfo {author} {\bibfnamefont
  {P.}~\bibnamefont {Lan}}, \ and\ \bibinfo {author} {\bibfnamefont
  {P.}~\bibnamefont {Lu}},\ }\bibfield  {title} {\enquote {\bibinfo {title}
  {High-order-harmonic generation of a doped semiconductor},}\ }\href {\doibase
  10.1103/PhysRevA.96.043425} {\bibfield  {journal} {\bibinfo  {journal} {Phys.
  Rev. A}\ }\textbf {\bibinfo {volume} {96}},\ \bibinfo {pages} {043425}
  (\bibinfo {year} {2017})}\BibitemShut {NoStop}%
\bibitem [{\citenamefont {Almalki}\ \emph {et~al.}(2018)\citenamefont
  {Almalki}, \citenamefont {Parks}, \citenamefont {Bart}, \citenamefont
  {Corkum}, \citenamefont {Brabec},\ and\ \citenamefont
  {McDonald}}]{Almalki2018}%
  \BibitemOpen
  \bibfield  {author} {\bibinfo {author} {\bibfnamefont {S.}~\bibnamefont
  {Almalki}}, \bibinfo {author} {\bibfnamefont {A.~M.}\ \bibnamefont {Parks}},
  \bibinfo {author} {\bibfnamefont {G.}~\bibnamefont {Bart}}, \bibinfo {author}
  {\bibfnamefont {P.~B.}\ \bibnamefont {Corkum}}, \bibinfo {author}
  {\bibfnamefont {T.}~\bibnamefont {Brabec}}, \ and\ \bibinfo {author}
  {\bibfnamefont {C.~R.}\ \bibnamefont {McDonald}},\ }\bibfield  {title}
  {\enquote {\bibinfo {title} {High harmonic generation tomography of
  impurities in solids: Conceptual analysis},}\ }\href {\doibase
  10.1103/PhysRevB.98.144307} {\bibfield  {journal} {\bibinfo  {journal} {Phys.
  Rev. B}\ }\textbf {\bibinfo {volume} {98}},\ \bibinfo {pages} {144307}
  (\bibinfo {year} {2018})}\BibitemShut {NoStop}%
\bibitem [{\citenamefont {Chinzei}\ and\ \citenamefont
  {Ikeda}(2020)}]{Chinzei2020}%
  \BibitemOpen
  \bibfield  {author} {\bibinfo {author} {\bibfnamefont {K.}~\bibnamefont
  {Chinzei}}\ and\ \bibinfo {author} {\bibfnamefont {T.~N.}\ \bibnamefont
  {Ikeda}},\ }\bibfield  {title} {\enquote {\bibinfo {title} {Disorder effects
  on the origin of high-order harmonic generation in solids},}\ }\href
  {\doibase 10.1103/PhysRevResearch.2.013033} {\bibfield  {journal} {\bibinfo
  {journal} {Phys. Rev. Research}\ }\textbf {\bibinfo {volume} {2}},\ \bibinfo
  {pages} {013033} (\bibinfo {year} {2020})}\BibitemShut {NoStop}%
\bibitem [{\citenamefont {\'{O}scar Zurr\'{o}n-Cifuentes}\ \emph
  {et~al.}(2019)\citenamefont {\'{O}scar Zurr\'{o}n-Cifuentes}, \citenamefont
  {Boyero-Garc\'{i}a}, \citenamefont {Hern\'{a}ndez-Garc\'{i}a}, \citenamefont
  {Pic\'{o}n},\ and\ \citenamefont {Plaja}}]{Zurron-Cifuentes2019}%
  \BibitemOpen
  \bibfield  {author} {\bibinfo {author} {\bibnamefont {\'{O}scar
  Zurr\'{o}n-Cifuentes}}, \bibinfo {author} {\bibfnamefont {R.}~\bibnamefont
  {Boyero-Garc\'{i}a}}, \bibinfo {author} {\bibfnamefont {C.}~\bibnamefont
  {Hern\'{a}ndez-Garc\'{i}a}}, \bibinfo {author} {\bibfnamefont
  {A.}~\bibnamefont {Pic\'{o}n}}, \ and\ \bibinfo {author} {\bibfnamefont
  {L.}~\bibnamefont {Plaja}},\ }\bibfield  {title} {\enquote {\bibinfo {title}
  {Optical anisotropy of non-perturbative high-order harmonic generation in
  gapless graphene},}\ }\href {\doibase 10.1364/OE.27.007776} {\bibfield
  {journal} {\bibinfo  {journal} {Opt. Express}\ }\textbf {\bibinfo {volume}
  {27}},\ \bibinfo {pages} {7776--7786} (\bibinfo {year} {2019})}\BibitemShut
  {NoStop}%
\bibitem [{\citenamefont {Zhang}\ \emph {et~al.}(2019)\citenamefont {Zhang},
  \citenamefont {Li}, \citenamefont {Zhou}, \citenamefont {Yue}, \citenamefont
  {Du}, \citenamefont {Fu},\ and\ \citenamefont {Luo}}]{Zhang2019}%
  \BibitemOpen
  \bibfield  {author} {\bibinfo {author} {\bibfnamefont {X.}~\bibnamefont
  {Zhang}}, \bibinfo {author} {\bibfnamefont {J.}~\bibnamefont {Li}}, \bibinfo
  {author} {\bibfnamefont {Z.}~\bibnamefont {Zhou}}, \bibinfo {author}
  {\bibfnamefont {S.}~\bibnamefont {Yue}}, \bibinfo {author} {\bibfnamefont
  {H.}~\bibnamefont {Du}}, \bibinfo {author} {\bibfnamefont {L.}~\bibnamefont
  {Fu}}, \ and\ \bibinfo {author} {\bibfnamefont {H.-G.}\ \bibnamefont {Luo}},\
  }\bibfield  {title} {\enquote {\bibinfo {title} {Ellipticity dependence
  transition induced by dynamical bloch oscillations},}\ }\href {\doibase
  10.1103/PhysRevB.99.014304} {\bibfield  {journal} {\bibinfo  {journal} {Phys.
  Rev. B}\ }\textbf {\bibinfo {volume} {99}},\ \bibinfo {pages} {014304}
  (\bibinfo {year} {2019})}\BibitemShut {NoStop}%
\bibitem [{\citenamefont {Xiao}\ \emph {et~al.}(2010)\citenamefont {Xiao},
  \citenamefont {Chang},\ and\ \citenamefont {Niu}}]{Xiao2010}%
  \BibitemOpen
  \bibfield  {author} {\bibinfo {author} {\bibfnamefont {D.}~\bibnamefont
  {Xiao}}, \bibinfo {author} {\bibfnamefont {M.-C.}\ \bibnamefont {Chang}}, \
  and\ \bibinfo {author} {\bibfnamefont {Q.}~\bibnamefont {Niu}},\ }\bibfield
  {title} {\enquote {\bibinfo {title} {Berry phase effects on electronic
  properties},}\ }\href {\doibase 10.1103/RevModPhys.82.1959} {\bibfield
  {journal} {\bibinfo  {journal} {Rev. Mod. Phys.}\ }\textbf {\bibinfo {volume}
  {82}},\ \bibinfo {pages} {1959} (\bibinfo {year} {2010})}\BibitemShut
  {NoStop}%
\bibitem [{\citenamefont {Zaks}\ \emph {et~al.}(2012)\citenamefont {Zaks},
  \citenamefont {Liu},\ and\ \citenamefont {Sherwin}}]{Zaks2012}%
  \BibitemOpen
  \bibfield  {author} {\bibinfo {author} {\bibfnamefont {B.}~\bibnamefont
  {Zaks}}, \bibinfo {author} {\bibfnamefont {R.~B.}\ \bibnamefont {Liu}}, \
  and\ \bibinfo {author} {\bibfnamefont {M.~S.}\ \bibnamefont {Sherwin}},\
  }\bibfield  {title} {\enquote {\bibinfo {title} {Experimental observation of
  electron-hole recollisions},}\ }\href {\doibase 10.1038/nature10864}
  {\bibfield  {journal} {\bibinfo  {journal} {Nature}\ }\textbf {\bibinfo
  {volume} {483}},\ \bibinfo {pages} {580} (\bibinfo {year}
  {2012})}\BibitemShut {NoStop}%
\bibitem [{\citenamefont {Langer}\ \emph {et~al.}(2016)\citenamefont {Langer},
  \citenamefont {Hohenleutner}, \citenamefont {Schmid}, \citenamefont
  {Poellmann}, \citenamefont {Nagler}, \citenamefont {Korn}, \citenamefont
  {Sch{\"u}ller}, \citenamefont {Sherwin}, \citenamefont {Huttner},
  \citenamefont {Steiner}, \citenamefont {Koch}, \citenamefont {Kira},\ and\
  \citenamefont {Huber}}]{Langer2016}%
  \BibitemOpen
  \bibfield  {author} {\bibinfo {author} {\bibfnamefont {F.}~\bibnamefont
  {Langer}}, \bibinfo {author} {\bibfnamefont {M.}~\bibnamefont
  {Hohenleutner}}, \bibinfo {author} {\bibfnamefont {C.~P.}\ \bibnamefont
  {Schmid}}, \bibinfo {author} {\bibfnamefont {C.}~\bibnamefont {Poellmann}},
  \bibinfo {author} {\bibfnamefont {P.}~\bibnamefont {Nagler}}, \bibinfo
  {author} {\bibfnamefont {T.}~\bibnamefont {Korn}}, \bibinfo {author}
  {\bibfnamefont {C.}~\bibnamefont {Sch{\"u}ller}}, \bibinfo {author}
  {\bibfnamefont {M.~S.}\ \bibnamefont {Sherwin}}, \bibinfo {author}
  {\bibfnamefont {U.}~\bibnamefont {Huttner}}, \bibinfo {author} {\bibfnamefont
  {J.~T.}\ \bibnamefont {Steiner}}, \bibinfo {author} {\bibfnamefont {S.~W.}\
  \bibnamefont {Koch}}, \bibinfo {author} {\bibfnamefont {M.}~\bibnamefont
  {Kira}}, \ and\ \bibinfo {author} {\bibfnamefont {R.}~\bibnamefont {Huber}},\
  }\bibfield  {title} {\enquote {\bibinfo {title} {Lightwave-driven
  quasiparticle collisions on a subcycle timescale},}\ }\href
  {https://doi.org/10.1038/nature17958} {\bibfield  {journal} {\bibinfo
  {journal} {Nature}\ }\textbf {\bibinfo {volume} {533}},\ \bibinfo {pages}
  {225 EP --} (\bibinfo {year} {2016})}\BibitemShut {NoStop}%
\bibitem [{\citenamefont {Banks}\ \emph {et~al.}(2017)\citenamefont {Banks},
  \citenamefont {Wu}, \citenamefont {Valovcin}, \citenamefont {Mack},
  \citenamefont {Gossard}, \citenamefont {Pfeiffer}, \citenamefont {Liu},\ and\
  \citenamefont {Sherwin}}]{Banks2017}%
  \BibitemOpen
  \bibfield  {author} {\bibinfo {author} {\bibfnamefont {H.~B.}\ \bibnamefont
  {Banks}}, \bibinfo {author} {\bibfnamefont {Q.}~\bibnamefont {Wu}}, \bibinfo
  {author} {\bibfnamefont {D.~C.}\ \bibnamefont {Valovcin}}, \bibinfo {author}
  {\bibfnamefont {S.}~\bibnamefont {Mack}}, \bibinfo {author} {\bibfnamefont
  {A.~C.}\ \bibnamefont {Gossard}}, \bibinfo {author} {\bibfnamefont
  {L.}~\bibnamefont {Pfeiffer}}, \bibinfo {author} {\bibfnamefont {R.-B.}\
  \bibnamefont {Liu}}, \ and\ \bibinfo {author} {\bibfnamefont {M.~S.}\
  \bibnamefont {Sherwin}},\ }\bibfield  {title} {\enquote {\bibinfo {title}
  {Dynamical birefringence: Electron-hole recollisions as probes of berry
  curvature},}\ }\href {\doibase 10.1103/PhysRevX.7.041042} {\bibfield
  {journal} {\bibinfo  {journal} {Phys. Rev. X}\ }\textbf {\bibinfo {volume}
  {7}},\ \bibinfo {pages} {041042} (\bibinfo {year} {2017})}\BibitemShut
  {NoStop}%
\bibitem [{\citenamefont {Langer}\ \emph {et~al.}(2018)\citenamefont {Langer},
  \citenamefont {Schmid}, \citenamefont {Schlauderer}, \citenamefont {Gmitra},
  \citenamefont {Fabian}, \citenamefont {Nagler}, \citenamefont {Sch{\"u}ller},
  \citenamefont {Korn}, \citenamefont {Hawkins}, \citenamefont {Steiner},
  \citenamefont {Huttner}, \citenamefont {Koch}, \citenamefont {Kira},\ and\
  \citenamefont {Huber}}]{Langer2018}%
  \BibitemOpen
  \bibfield  {author} {\bibinfo {author} {\bibfnamefont {F.}~\bibnamefont
  {Langer}}, \bibinfo {author} {\bibfnamefont {C.~P.}\ \bibnamefont {Schmid}},
  \bibinfo {author} {\bibfnamefont {S.}~\bibnamefont {Schlauderer}}, \bibinfo
  {author} {\bibfnamefont {M.}~\bibnamefont {Gmitra}}, \bibinfo {author}
  {\bibfnamefont {J.}~\bibnamefont {Fabian}}, \bibinfo {author} {\bibfnamefont
  {P.}~\bibnamefont {Nagler}}, \bibinfo {author} {\bibfnamefont
  {C.}~\bibnamefont {Sch{\"u}ller}}, \bibinfo {author} {\bibfnamefont
  {T.}~\bibnamefont {Korn}}, \bibinfo {author} {\bibfnamefont {P.~G.}\
  \bibnamefont {Hawkins}}, \bibinfo {author} {\bibfnamefont {J.~T.}\
  \bibnamefont {Steiner}}, \bibinfo {author} {\bibfnamefont {U.}~\bibnamefont
  {Huttner}}, \bibinfo {author} {\bibfnamefont {S.~W.}\ \bibnamefont {Koch}},
  \bibinfo {author} {\bibfnamefont {M.}~\bibnamefont {Kira}}, \ and\ \bibinfo
  {author} {\bibfnamefont {R.}~\bibnamefont {Huber}},\ }\bibfield  {title}
  {\enquote {\bibinfo {title} {Lightwave valleytronics in a monolayer of
  tungsten diselenide},}\ }\href {\doibase 10.1038/s41586-018-0013-6}
  {\bibfield  {journal} {\bibinfo  {journal} {Nature}\ }\textbf {\bibinfo
  {volume} {557}},\ \bibinfo {pages} {76--80} (\bibinfo {year}
  {2018})}\BibitemShut {NoStop}%
\bibitem [{\citenamefont {Li}\ \emph {et~al.}(2019{\natexlab{b}})\citenamefont
  {Li}, \citenamefont {Zhang}, \citenamefont {Fu}, \citenamefont {Feng},
  \citenamefont {Hu},\ and\ \citenamefont {Du}}]{Li2019phase}%
  \BibitemOpen
  \bibfield  {author} {\bibinfo {author} {\bibfnamefont {J.}~\bibnamefont
  {Li}}, \bibinfo {author} {\bibfnamefont {X.}~\bibnamefont {Zhang}}, \bibinfo
  {author} {\bibfnamefont {S.}~\bibnamefont {Fu}}, \bibinfo {author}
  {\bibfnamefont {Y.}~\bibnamefont {Feng}}, \bibinfo {author} {\bibfnamefont
  {B.}~\bibnamefont {Hu}}, \ and\ \bibinfo {author} {\bibfnamefont
  {H.}~\bibnamefont {Du}},\ }\bibfield  {title} {\enquote {\bibinfo {title}
  {Phase invariance of the semiconductor bloch equations},}\ }\href {\doibase
  10.1103/PhysRevA.100.043404} {\bibfield  {journal} {\bibinfo  {journal}
  {Phys. Rev. A}\ }\textbf {\bibinfo {volume} {100}},\ \bibinfo {pages}
  {043404} (\bibinfo {year} {2019}{\natexlab{b}})}\BibitemShut {NoStop}%
\end{thebibliography}

%

\end{document}